\documentstyle[12pt]{article}

\newcommand{\e}{\varepsilon}
\newcommand{\p}{\bot}
\newcommand{\pa}{\scriptscriptstyle \|}
\newcommand{\dd}{\partial}
\newcommand{\ka}{\char'32}
\newcommand{\de}{\delta}
\newcommand{\om}{\omega}
\newcommand{\Om}{\Omega}
\newcommand{\f}{\varphi}
\newcommand{\ls}{\left(}
\newcommand{\lks}{\left[}
\newcommand{\rs}{\right)}
\newcommand{\rks}{\right]}
\newcommand{\g}{\gamma}
\newcommand{\al}{\alpha}
\newcommand{\bet}{\beta}
\newcommand{\ta}{\tau}
\newcommand{\n}{\nu}
\newcommand{\m}{\mu}
\newcommand{\s}{\sigma}
\newcommand{\La}{\Lambda}
\newcommand{\la}{\lambda}
\newcommand{\et}{\eta}
\newcommand{\te}{\theta}
\newcommand{\ps}{\psi}
\newcommand{\ra}{\rangle}
\newcommand{\ilm}{\int\limits}
\newcommand{\str}[1]{\mathrel{\mathop{\longrightarrow}\limits_{#1}}}

\newcommand{\st}{\protect\\}

\newcommand{\disn}[2]{$$\displaylines{\refstepcounter{equation}%
            \label{#1}\hskip 1em minus 1em #2\hfilneg}$$}
\newcommand{\nom}{\hfil\hskip 1em minus 1em (\theequation)}
\newcommand{\no}{\hfil \hskip 1em minus 1em\phantom{(\theequation)}%
            \hfilneg\cr\hfilneg\hskip 1em minus 1em\hfil}
\newcommand{\nol}{\hfill\no}
\newcommand{\nor}{\no\hfill}
\newcommand{\nolr}{\hfill\no\hfill}
\newcommand{\nop}{\hfill\cr\hfil}

\makeatletter

\newcount\dobav
\newcount\otlog
\newcount\vrem
\def\@citex[#1]#2{\if@filesw\immediate\write\@auxout{\string\citation{#2}}\fi
  \let\@citea\@empty
  \dobav=-1
  \otlog=-1
  \@cite{\@for\@citeb:=#2\do
    {\def\@tempa##1##2\@nil{\edef\@citeb{\if##1\space##2\else##1##2\fi}}%
     \expandafter\@tempa\@citeb\@nil
     \@ifundefined{b@\@citeb}{\@warning%
       {Citation `\@citeb' on page \thepage \space undefined}%
       \vrem=-1}{\vrem=\csname b@\@citeb\endcsname}
\advance\vrem by -1 \ifnum \vrem=\dobav
 \otlog=\vrem
 \advance\otlog by 1
\else
 \ifnum \vrem=\otlog
  \advance\otlog by 1
 \else
  \ifnum \otlog>0
   \advance\dobav by 1
   \ifnum \otlog=\dobav
    \hbox{,\penalty\@m\ \the\otlog}%
   \else
    \hbox{--\the\otlog}%
   \fi
   \otlog=-1
  \fi
  \dobav=\vrem
  \advance\dobav by 1
  \@citea\def\@citea{,\penalty\@m\ }%
  \ifnum \dobav=-1
   {\reset@font\bf ?}%
  \else
   \hbox{\the\dobav}%
  \fi
 \fi
\fi
}%
\ifnum \otlog>0
 \advance\dobav by 1
 \ifnum \otlog=\dobav
  \hbox{,\penalty\@m\ \the\otlog}%
 \else
  \hbox{--\the\otlog}%
 \fi
\fi }{#1}}

\long\def\@makecaption#1#2{%
   \vskip 10\p@
   \setbox\@tempboxa\hbox{#1. #2}%
   \ifdim \wd\@tempboxa >\hsize
       #1. #2\par
     \else
       \hbox to\hsize{\hfil\box\@tempboxa\hfil}%
   \fi}

\def\newpic#1{%
   \def\emline##1##2##3##4##5##6{%
      \put(##1,##2){\special{em:point #1##3}}%
      \put(##4,##5){\special{em:point #1##6}}%
      \special{em:line #1##3,#1##6}}}
\newpic{}

\makeatother

\begin{document}

\title{Quantization of Field Theory on the Light Front}

\author{V.~A.~Franke, Yu.~V.~Novozhilov, S.~A.~Paston, E.~V.~Prokhvatilov\\
        {\it St.-Petersburg State University, Russia}}

\date{\vskip 15mm}

\maketitle

\begin{abstract}
\normalsize
Canonical formulation of quantum field theory on the Light Front (LF)
is reviewed. The problem of constructing the LF Hamiltonian which gives
the theory equivalent to original  Lorentz and gauge invariant one is
considered. We describe possible ways of solving this problem: (a) the
limiting transition from the equal-time Hamiltonian in a fast moving
Lorentz frame to LF Hamiltonian, (b) the direct comparison of LF
perturbation theory in coupling constant and usual Lorentz-covariant
Feynman perturbation theory. The results of the application of method (b)
to QED-1+1 and QCD-3+1 are given. Gauge invariant regularization
of LF Hamiltonian via introducing a lattice in transverse coordinates
and imposing periodic boundary conditions in LF coordinate $x^-$
for gauge fields on the interval $|x^-|<L$ is also considered.
\end{abstract}

\newpage
\section{Introduction}
A possible approach to solving the strong interaction field
theory is the canonical quantization on the Light-Front (LF)
with the application of corresponding Schroedinger equation.
To realize such program one introduces LF coordinates \cite{dirak}
$x^{\pm}=(x^0\pm x^3)/\sqrt{2}, x^1, x^2,$
where $x^0, x^1, x^2, x^3$ are Lorentz coordinates. The $x^+$ plays the
role of time, and canonical quantization is carried out on a hypersurface
$x^+=const$.
The advantage of this scheme is connected with the positivity of the
momentum $P_-$ (translation operator along $x^-$ axis), which becomes
quadratic in fields on the LF. As a consequence the lowest eigenstate
of the operator $P_-$ is both physical vacuum and the "mathematical" vacuum
of perturbation theory \cite{leut}. Using Fock space \cite{fock}
over this vacuum one can solve
stationary Schroedinger equation with Hamiltonian $P_+$ (translation operator
along $x^+$ axis) to find the spectrum of bound states. The problem of
describing the physical vacuum, very complicated in usual formulation
with Lorentz coordinates, does not appear here.
Such approach is called LF Hamiltonian approach.
It attracts attention for a long time as a possible mean for solving
Quantum Field Theory problems.

  While giving essential advantages, the application of LF coordinates in
Quantum Field Theory  leads to some difficulties. The hyperplane
$x^+=\mbox{const}$
 is a characteristic  surface for relativistic differential field equations.
It is not evident without additional investigation that
quantization  on such hypersurface generates a theory equivalent to one
quantized in the usual way in Lorentz coordinates
\cite{chang1,chang2,yan,pred1,pred2,pirner,burk1,tmf97,lenc}.
This is in particular
essential because of the special divergences at $p_-=0$ appearing in LF
quantization scheme. Beside of usual ultraviolet regularization one has to
apply special regularization of such divergences. We will consider the
following simplest prescription of such regularization:

(a) cutoff of momenta $p_-$
\disn{1.1}{|p_- |\ge\e ,\quad\e>0;\nom}

(b) cutoff of the $x^-$
\disn{1.2}{-L\le x^-\le L.\nom}
with periodic boundary conditions in $x^-$ for all fields.

The regularization (b) discretizes the spectrum of
the operator $P_-$ ($p_-=\pi n/L$, where $n$ is an integer). This formulation
is called sometime "Discretized LF Quantization" \cite{brodsk}.
Fourier components of
fields, corresponding to $p_-=0$ (and usually called "zero modes")  turn out
to be dependent variables and must be expressed in terms of nonzero modes  via
solving constraint equations (constraints). These constraints are usually very complicated,
and solving of them is a difficult problem.

The prescriptions of regularization of divergences at $p_-=0$
described above are
convenient for Hamiltonian approach,
 but both of them break Lorentz
invariance and the prescription (a) breaks  also the gauge invariance.
Therefore the
equivalence of LF and original Lorentz (and gauge) invariant formulation
can be broken even in the limit of removed cutoff. To avoid  this
nonequivalence  some modification of usual renormalization procedure may be
necessary, see for example \cite{tmf97,tmf99} and \cite{kniga}.
The problem of constructing the LF Hamiltonian which gives a theory
equivalent to original  Lorentz and gauge invariant one turned out to be
rather difficult. We will describe possible approaches to this problem.

In sect.~2 we give basic relations of quantum field theory in LF
coordinates. In sect.~3 we consider the limiting transition from fast
moving Lorentz frame to the LF. This transition relates
Hamiltonian formulations in Lorentz and LF coordinates
and firstly was presented in \cite{pred1,pred2}.
In sect.~4 we investigate the
relation between LF perturbation theory in coupling constant and usual
Lorentz-covariant Feynman perturbation theory.
With the help of this investigation we show how to construct
LF Hamiltonian giving a theory
perturbatively equivalent to original one for
Yukawa model, for QCD in four-dimensional space-time,
and  for QED in two-dimensional space-time
(originally this was considered in \cite{tmf97,tmf99,shw2,tmf02,rasch}.
In sect.~5 we carry out  gauge
invariant ultraviolet regularization of LF Hamiltonian introducing a
lattice in transverse coordinates  $x^1, x^2$  and using,
instead of transverse components of gauge fields,
complex matrices in color space as independent gauge
variables on the links of the lattice \cite{heplat,tmf04}.
We find that LF canonical
formalism for gauge theories with this regularization  avoid usual most
complicated constraints connecting zero and nonzero modes.
However, the quantization leads to Lorentz-noninvariant results.

\section{Formal canonical quantization of Field Theory \st on the Light Front
     and the problem of bound states}

    In order to find the bound state spectrum in some field theory
quantized on the LF the following system of equations is usually solved:
 \disn{2.1a}{
P_+ |\Psi\rangle=P'_+ |\Psi\rangle,
\nom}
\vskip -8mm
 \disn{2.1b}{
P_- |\Psi\rangle=P'_- |\Psi\rangle,
\nom}
\vskip -8mm
 \disn{2.1v}{P_\p |\Psi\rangle =0,\nom}
where $P_\bot =\{P_1,P_2\}$. The mass of bound state is equal to
\disn{2.2}{m = \sqrt{2P'_+P'_-}.\nom}

It was taken into account that nonzero components of metric tensor in LF
coordinates are
\disn{2.3}{
g_{-+}=g_{+-}=1,\quad g_{11}=g_{22}=-1.
\nom}
The operators $P_-$, $P_\bot $ are quadratic in fields, and the solution of
equations (\ref{2.1b}), (\ref{2.1v})
is trivial. The problem is in solving the
Schroedinger equation (\ref{2.1a}).
Physical vacuum $|\Omega\rangle $ is lowest eigenstate of the operator
$P_- $, and must satisfy the equations:
\disn{2.4a}{
P_+ |\Omega\rangle =0,
\nom}
\vskip -8mm
\disn{2.4b}{
P_- |\Omega\rangle =0,
\nom}
\vskip -8mm
\disn{2.4v}{
P_\p |\Omega\rangle =0.
\nom}
To fulfil equations (\ref{2.4a}), (\ref{2.4b}) one should subtract , if it is
necessary, corresponding renormalizing constants from the
operators $P_+$, $P_-$.
The $|\Omega\rangle $ plays simultaneously the role of mathematical vacuum of
LF Fock space. A solution $|\Psi\rangle$ of Schrodinger equation (\ref{2.1a})
belongs to this space.

Expressions for the operators $P_+$, $P_-$ can be obtained by canonical
quantization on the LF. Let us describe the procedure of such quantization
via some examples, without analyzing so far the question about
the equivalence of appearing theory and original Lorentz-covariant one.
It is assumed that in addition to explicit regularization of divergences at
$p_-=0$ also ultraviolet regularization is implied.

\subsection{Scalar selfinteracting field in  (1+1)-dimensional space-time.}
Peculiarities of LF quantization are well seen even in this simple
example.  We have only LF coordinates $x^+$, $x^-$. The Lagrangian is equal to
\disn{2.5}{
L=\int dx^-\,\ls\dd _\mu\f\dd ^\mu\f -
\frac{1}{2}m^2\f^2-\la\f^4\rs,
\nom}
or
\disn{2.6}{
L=\int dx^-\,\ls\frac{1}{2}\dd _+\f\dd _-\f -
\frac{1}{2}m^2\f^2-\la\f^4\rs,
\nom}
The "time" derivative $\dd_+\f$ enters into this Lagrangian only linearly.
For the transition to canonical theory it is sufficient to rewrite the
expression $\int dx^-\,\frac{1}{2}\dd _+\f\dd _-\f$ in standard canonical
form. To achieve this let us take the Fourier decomposition
\disn{2.7}{
\f(x^-)=(2\pi)^{-\frac{1}{2}}\ilm_0^\infty dk\,|2k|^{-\frac{1}{2}}
\ls a(k)\exp(-ikx^-)+a^+(k)\exp(ikx^-)\rs,
\nom}
where $k\equiv k_-,$ $\f(x^-)\equiv\f (x^+,x^-),$ $a(k)\equiv a(x^+,k)$.
The Lagrangian (\ref{2.6}) takes the form
\disn{2.8}{
L=\ilm_0^\infty dk\,\ls\frac{a^+(k)\dot a(k)-a(k)\dot  a^+(k)}{2i}
\rs -H,
\nom}
where $\dot a\equiv\dd a/\dd x^+$ and
\disn{2.9}{
H=\int dx^-\,\ls\frac{1}{2}m^2\f^2+\la\f^4\rs.
\nom}
Here we have used the equality
\disn{2.10}{
\ilm_0^\infty dk\,\ilm_0^\infty dk'\,\de (k+k')k'\ls
a(k)\dot a(k')-a^+(k)\dot  a^+(k')\rs =0.
\nom}
It is implied that the function $\f (x)$ in (\ref{2.9}) is expressed in terms of
$a^+(k)$ and $a(k)$ with the help of formulae (\ref{2.6}). "Time" derivatives
$\dot a(k),$ $\dot a^+(k)$ enter into Lagrangian $L$ in a form standard for
canonical theory. Therefore one can interpret after quantization
the $a^+(k)$
and $a(k)$ as creation and annihilation operators satisfying the following
commutation relations at fixed $x^+$ and $k>0,$ $k'>0$:
\disn{2.11}{
[a(k),a^+(k')]=\de (k-k'),\quad [a(k),a(k')]=0,
\nom}
It is also seen that the  $H$ is LF Hamiltonian, i.e. $H=P_+$.

   We have also the formulae
\disn{2.12}{P_\mu =\int dx^-\,T_{-\mu},\nom}
where the energy-momentum tensor $T_{\nu\mu}$ is equal to
\disn{2.13}{T_{\nu\mu}=\dd_\nu\f\dd_\mu\f-g_{\mu\nu}{\cal L}.\nom}
Via this relation one can reproduce the expression
(\ref{2.9}) for $P_+\equiv H$ , and obtain
the equality
\disn{2.14}{P_-=\int dx^-\,(\dd_-\f)^2=\frac{1}{2}\ilm_0^\infty dk\, k\ls
a^+(k)a(k)+a(k)a^+(k)\rs.\nom}

The lowest eigenstate of the operator $P_-$ is the physical vacuum
$|\Omega\rangle$ for which
\disn{2.15}{a(k)|\Omega\rangle =0\nom}
at any $k$. It is seen that vacuum expectation values $\langle\Omega|P_-|
\Omega\rangle$,
$\langle\Omega|P_+|\Omega\rangle$ are
infinite. The renormalization  can be got by taking normal ordered forms
$:\!P_+\!:$, $:\!P_-\!:$ with respect to the operators $a^+$, $a$ (the symbol
$::$ means as usual that operators $a^+$ stand everywhere before of operators
$a$). Normal ordering of the $\lambda\f^4$ term in the Hamiltonian $P_+$
leads also to renormalization of the mass. In the following, writing $P_+$,
$P_-$, we mean the expressions $:\!P_+\!:$, $:\!P_-\!:$, satisfying
conditions (\ref{2.4a}), (\ref{2.4b}). Normal ordering of the operators
$P_+$, $P_-$  allows to avoid all ultraviolet divergences in this simple model.

\subsection{Theory of interacting scalar and fermion fields\st
in (3+1)-dimensional space-time (Yukawa model)}

The Lagrangian of the model is
\disn{2.16}{L=\int d^2x^\bot dx^-\ls\overline{\psi}\ls i\gamma^\mu\dd_\mu-M
\rs\psi +\frac{1}{2}\dd_\mu\f\dd^\mu\f-\frac{1}{2}m^2\f^2-g
\overline{\psi}\psi \f-\la'\f^3-\la\f^4\rs,\nom}
where $M$ is the fermion mass, $m$ is the boson mass, $\overline{\psi}=\psi^+
\gamma^0$, $\f=\f^+$; $g$, $\la$, $\la'$ are coupling constants.
Here and so on $\mu,\nu,\ldots=+,-,1,2$; $i,k,\ldots = 1,2$; $x^\bot\equiv\ls
x^1,x^2\rs.$
For Dirac's $\gamma$-matrices we use:
\disn{2.17}{\gamma^0 = \ls\begin{array}{cc}0 & -iI\\iI & 0\end{array}\rs,\qquad
\gamma^3 =\ls\begin{array}{cc}0 & iI\\iI & 0\end{array}\rs,\qquad
\gamma^\bot =\ls\begin{array}{cc}-i\sigma^\bot & 0\\0 & i\sigma^\bot
\end{array}\rs,\nom}
where $I$ is a unit $2\times 2$ matrix, $\sigma^\bot\equiv\{\sigma^1,\sigma^2
\}$, $\sigma^i$ are Pauli matrices.

We introduce $2$-component spinors $\chi$, $\xi$, writing
\disn{nnn1}
{\psi=\ls
\begin{array}{c}\chi\\\xi\end{array}\rs.
\nom}
The Lagrangian $L$ can be written in the form
\disn{2.18}{
L=\int d^2x^\bot dx^-\,\Bigl( i\sqrt{2}\chi^+\dd_+\chi+
i\sqrt{2}\xi^+\dd_-\xi+\ls i\xi^+\ls \sigma^i\dd_i-M\rs\chi+\mbox{h.c.}
\rs +\nolr
+\dd_+\f\dd_-\f-\frac{1}{2}\dd_i\f\dd_i\f
-\frac{1}{2}m^2\f^2-ig\f\ls \xi^+\chi-\chi^+\xi\rs -\la'\f^3-\la\f^4
\Bigr),
\nom}
where h.c. means Hermitian conjugation.
The variation of this Lagrangian with respect to $\chi^+$
leads to the equation
\disn{2.19}{\sqrt{2}\dd_-\xi=-\ls\s^i\dd_i-M\rs\chi+g\f\chi.\nom}
This equation does not contain derivatives in $x^+$  and therefore is a
constraint. One should solve it with respect to $\xi$ and substitute the
result into the Lagrangian. In doing this we must invert the operator
$\dd_-$ which becomes an operator of multiplication  $ik_-$ after
Fourier transformation. Inverse operator $(ik_-)^{-1}$ has singularity at
$k_-=0$. To avoid this singularity we introduce the regularization (\ref{1.1}).
For any function $f(x^-)\equiv f(x^+,x^-,x^\bot)$ we define Fourier transform
\disn{2.20}{f(x^-)=\frac{1}{\sqrt{2\pi}}\int dk_-\,e^{ik_-x^-}\tilde f(k_-),\nom}
where $\tilde f(k_-)\equiv\tilde f(x^+,k_-,x^\bot)$, and put
\disn{2.21}{[f(x^-)]\equiv\frac{1}{\sqrt{2\pi}}\int dk_-\,e^{ik_-x^-}\tilde f(k_-),
\qquad |k_-|\ge\e>0.\nom}

We insert into the Lagrangian (\ref{2.18}) the variables
$[\chi]$, $[\chi^+]$, $[\xi]$, $[\xi^+]$, $[\f]$   instead  of
$\chi$, $\chi^+$, $\xi$, $\xi^+$, $\f$
and obtain  the constraint equation
\disn{2.22}{\sqrt{2}\dd_-[\xi]=-\ls\s^i\dd_i-M\rs [\chi]+g\big[[\f][\chi]\big]
\nom}
instead of (\ref{2.19}).
Its solution is
\disn{2.23}{[\xi]=\frac{1}{\sqrt{2}}\dd^{-1}_-\Big(-\ls\s^i\dd_i-M\rs [\chi]+g\big[[\f]
[\chi]\big]\Big),\nom}
where the operator $\dd_-^{-1}$ is completely defined by the condition
\disn{2.24}{\dd_-^{-1}[f]=\lks\dd_-^{-1}[f]\rks.\nom}
After Fourier transformation the operator $\dd_-^{-1}$ is replaced by
$(ik_-)^{-1}$.

Substituting the expression (\ref{2.23}) into the Lagrangian (where all fields
$\chi$, $\chi^+,\ldots$ are replaced with $[\chi]$, $[\chi^+],\ldots$)
we come to the result:
\disn{2.25}{
L=\int d^2x^\bot dx^-\,\bigg( i\sqrt{2}\lks\chi^+\rks\dd_+\lks\chi\rks+
\dd_-[\f]\dd_+[\f]+\nol
+\frac{1}{\sqrt{2}}\Big(\ls\sigma^i\dd_i-M\rs[\chi]-
g\big[[\f][\chi]\big]\Big)^+\!\ls-i\dd_-\rs^{-1}\!\Big(\ls\sigma^k
\dd_k-M\rs[\chi]-g\big[ [\f][\chi]\big]\Big)-\nor
-\frac{1}{2}\dd_i[\f]\dd_i
[\f]-\frac{1}{2}m^2[\f]^2-\la'[\f]^3-\la[\f]^4\bigg),
\nom}
As in sect.~2a time derivatives $\dd_+[\chi]$, $\dd_+[\f]$ enter into the
Lagrangian (\ref{2.25}) linearly.  Therefore to go to canonical formalism
it is sufficient to find a standard canonical form for the expression
\disn{nnn2}{
i\sqrt{2}[\chi^+]\dd_+[\chi]+\dd_-[\f^+]\dd_+[\f]
\nom}
(before a quantization the quantities $\chi^+$, $\chi$ are elements of Grassman
algebra).  We write
\disn{2.26a}{\big[\f(x^-)\big]=(2\pi)^{-1/2}\ilm_\e^\infty dk_-\,(2k_-)^{-1/2}
\big(a(k_-)\exp(-ik_-x^-)+a^+(k_-)\exp(ik_-x^-)\big),\nom}
\vskip -8mm
\disn{2.26b}{\big[\chi_r(x^-)\big]=(2\pi)^{-1/2}2^{-1/4}\ilm_\e^\infty dk_-
\big(b_r(k_-)\exp(-ik_-x^-)+c^+_r(k_-)\exp(ik_-x^-)\big),\nom}
\vskip -8mm
\disn{2.26v}{\big[\chi_r^+(x^-)\big]=(2\pi)^{-1/2}2^{-1/4}\ilm_\e^\infty dk_-
\big(c_r(k_-)\exp(-ik_-x^-)+b^+_r(k_-)\exp(ik_-x^-)\big),\nom}
where
\disn{nnn3}{
\big[\f(x^-)\big]\equiv\big[\f(x^+,x^-,x^\bot)\big],\qquad
a(k_-)\equiv a(x^+,k_-,x^\bot)
\nom}
et cetera, $r=1,2$.
    The Lagrangian (\ref{2.25}) takes the form
\disn{2.27}{L=\int d^2x^\bot\ilm_\e^\infty dk_-\,\big(a(k_-)\dot a^+(k_-)
-a^+(k_-)\dot a(k_-)-\nolr
-b_r(k_-)\dot b_r^+(k_-)-b_r^+(k_-)\dot b_r(k_-)-
c_r(k_-)\dot c_r^+(k_-)-c_r^+(k_-)\dot c_r(k_-)\big)-H,\nom}
where
\disn{2.28}{H=\bigg(-\frac{1}{\sqrt{2}}\Big(\ls\sigma^i\dd_i-M\rs[\chi]-
g\big[[\f][\chi]\big]\Big)^+\!\ls-i\dd_-\rs^{-1}\!\Big(\ls\sigma^k
\dd_k-M\rs[\chi]-g\big[ [\f][\chi]\big]\Big)+\no+
\frac{1}{2}\dd_i[\f]\dd_i
[\f]+\frac{1}{2}m^2[\f]^2+\la'[\f]^3+\la[\f]^4\bigg).\nom}
It is assumed that the quantities $[\chi]$, $[\chi^+]$, $[\f]$ in the
formulae (\ref{2.28}) are expressed in terms of $b$, $b^+$, $c$, $c^+$, $a$,
$a^+$ with the help of (\ref{2.26a}), (\ref{2.26b}), (\ref{2.26v}).

It follows from (\ref{2.27}) that $a^+$, $a$, $b^+$, $b$, $c^+$, $c$
play a role of creation and annihilation operators. After quantization
they satisfy the commutation relations (at $x^+=$const):
\disn{2.29a}{[a(k_-,x^\bot),a^+(k'_-,x'^\bot)]_-=\de(k_--k'_-)\de^2(x^\bot-x'^\bot),\nom}
\vspace{-8mm}
\disn{2.29b}{[b(k_-,x^\bot),b^+(k'_-,x'^\bot)]_+=\de(k_--k'_-)\de^2(x^\bot-x'^\bot),\nom}
\vspace{-8mm}
\disn{2.29v}{[c(k_-,x^\bot),c^+(k'_-,x'^\bot)]_+=\de(k_--k'_-)\de^2(x^\bot-x'^\bot),\nom}
where $[x,y]_\pm=xy\pm yx$. The remaining (anti)commutators are equal to zero.
It is seen that the quantity $H$ is LF Hamiltonian ($H=P_+$). The operator
of the momentum $P_-$ is equal to
\disn{2.30}{
P_-=\int d^2xdx^-\,T_{--}=\frac{1}{2}\int d^2x^\bot\ilm_\e^\infty
dk_-\,k_-\big(a^+(k_-)a(k_-)
+a(k_-)a^+(k_-)+\nolr
+b_r^+(k_-)b_r(k_-)-b_r(k_-)b_r^+(k_-)+
c_r^+(k_-)c_r(k_-)-c_r(k_-)c_r^+(k_-)\big).
\nom}

The quantities $P_+\equiv H$ and $P_-$ should be normally ordered with
respect to creation and annihilation operators. The lowest eigenstate of
the momentum $P_-$ is the physical vacuum. It is defined by conditions
\disn{2.31}{
a(k_-,x^\bot)|\Omega\rangle =0,\quad
b(k_-,x^\bot)|\Omega\rangle =0,\quad
c(k_-,x^\bot)|\Omega\rangle =0,\quad\forall\;x^\bot ,\, k_-\ge\e>0.
\nom}
The equalities (\ref{2.4a}), (\ref{2.4b}) become true after normal ordering
of the operators $P_+$ and  $P_-$. For the $P_-$ it is seen from the formulae
(\ref{2.30}), and for the $P_+$ it
follows from the fact that every term of $P_+$ contains a $\de$-function
of difference between the sum of momenta $k_-$ of creation
operators and the sum of momenta $k_-$ of annihilation operators.
Due to the positivity of all momenta $k_-$, in our regularization scheme every
term of the $P_+$ contains at least one annihilation operator. Therefore for
normally ordered $P_+$ we  have $P_+|\Omega\rangle=0$.

The model under consideration requires ultraviolet regularization. It
can be done in different ways. We consider this question  together with the
renormalization problem in sect.~3 and 4.

\subsection{The U(N)-theory of pure gauge fields}
    We consider the $U(N)$ rather than the $SU(N)$
theory because it is technically
more simple. The transition to the $SU(N)$ can be done easily.
   Gauge field is described by Hermitian matrices
\disn{2.31a}{A_\mu(x)=A^+_\mu (x)\equiv\left\{A^{ij}_\mu(x)\right\},\nom}
where $\mu=+,-,1,2$;  $i,j=1,2,\ldots ,N$.
Let us assume, that for the indexes $i,j$ and analogous the usual rule
of summation on repeated indexes is not used, and where it is necessary
the sign of a sum is indicated.
   Field strengths tensor is
\disn{2.32}{F_{\mu\nu}=\dd_\mu A_\nu -\dd_\nu A_\mu -ig[A_\mu ,A_\nu],\nom}
and gauge transformation has the form
\disn{2.33}{A_\mu\to A'_\mu=U^+A_\mu U+\frac{i}{g}U^+\dd_\mu U,\quad U^+U=I.\nom}
   To escape a breakdown of gauge invariance we apply the regularization
of the type (\ref{1.2}) with periodic boundary conditions
\disn{2.34}{A_\mu(x^+,-L,x^\bot)=A_\mu(x^+,L,x^\bot),\nom}
on the interval $-L\le x^-\le L$
(this exact periodicity can be always achieved starting from
gauge invariant one, i.~e. the periodicity up to a gauge transformation).
All Fourier modes of $A_\mu(x)$
in $x^-$ must be kept, including zero modes (at $k_-=0$).

The Lagrangian has the form
\disn{2.35}{L=-\frac{1}{2}\int d^2x^\bot\ilm_{-L}^Ldx^-{\rm Tr}\ls F_{\mu\nu}
F^{\mu\nu}\rs,\nom}
or
\disn{2.36}{L=\int d^2x^\bot\ilm_{-L}^Ldx^-{\rm Tr}\ls F^2_{+-}+2F_{-k}F_{+k}-
\frac{1}{2}F_{kk'}F_{kk'}\rs,\nom}
where  $k,k'=1,2$. Time derivatives $\dd_+A_k$ are present only
in the term
\disn{nnn4}{
{\rm Tr}\ls 2(\dd_-A_k-\dd_kA_--ig[A_-,A_+])\dd_+A_k\rs.
\nom}
This expression can be rewritten in standard canonical form only after fixing
the gauge in a special form of the type $A_-=0$ because then the term above
becomes similar to that of scalar field theory
${\rm Tr}\ls 2(\dd_-A_k)\dd_+A_k\rs$.
However not every field, periodic in $x^-$, can be transformed to the
$A_-=0$ gauge.
Indeed, the loop integral
\disn{2.37}{\Gamma(x^+,x^\bot)={\rm Tr}\left\{{\rm P}\exp\ls i\ilm_{-L}^Ldx^-\,A_-
(x^+,x^-,x^\bot)\rs\right\},\nom}
where the symbol 'P' means the ordering of operators along the
$x^-$, is a
gauge invariant quantity. If this integral is not equal to $N$ for some field
then the gauge $A_-=0$ is not possible for this field. Therefore we
choose more weak gauge condition
\disn{2.38}{A^{ij}_-= 0,\quad i\ne j,\qquad \dd_-A^{ii}_-=0,\nom}
and put
\disn{2.39}{A_-^{ii}(x)=v^i(x^+,x^\bot).\nom}
The gauge (\ref{2.38}) breaks not only the local gauge invariance but,
unlike the $A_-=0$ gauge,
also global gauge invariance (it remains only the abelian subgroup of gauge
transformations not depending on $x^-$).  This has some technical
inconvenience  but now any periodic field can be described in the gauge
(\ref{2.38}). Furthermore, if we restrict the class of possible periodic fields
by the condition $A_-=0$, disregarding the described consideration,
we come to canonical theory with even more
complicate  constraints if zero modes are taken into account
\cite{nov1,nov2,nov3}.

From the point of view of LF canonical  formalism the variables $A_-^{ij}$
are "coordinates". Therefore one can restrict their values by the
condition (\ref{2.38}) directly in the Lagrangian without loosing any
equations of motion.
  Let us introduce the denotations
\disn{2.40a}{D_-A^{ij}_+=(\dd_--ig(v^i-v^j))A^{ij}_+,\nom}
\disn{2.40b}{D_-A^{ij}_k=(\dd_--ig(v^i-v^j))A^{ij}_k,\nom}
where obviously
\disn{2.40v}{D_-A^{ii}_k=\dd_-A^{ii}_k.\nom}
Also for any function $f(x^-)\equiv f(x^+,x^-,x^\bot)$ periodic in $x^-$
we denote
\disn{2.41a}{f_{(0)}=\frac{1}{2L}\ilm_{-L}^Ldx^-\,f(x^-),\nom}
\disn{2.41b}{[f(x^-)]=f(x^-)-f_{(0)}.\nom}
Obviously,
\disn{2.41v}{\ilm_{-L}^Ldx^-\,[f(x^-)]=0.\nom}

After gauge fixing (\ref{2.38}) the Lagrangian (\ref{2.36}) takes the form
\disn{2.42}
{L=\int d^2x\ilm_{-L}^Ldx^-\Bigg\{2\sum_i\ls\dd_-\lks A^{ii}_k\rks\rs
\dd_+\lks A^{ii}_k\rks +2\sum_{i,j,\, i\ne j}\ls D_-
A^{ij}_k\rs\dd^+A^{ji}_k+\sum_i\ls\dd_-\lks A^{ii}_+\rks\rs^2+\nop
+\!\!\!\sum_{i,j,\, i\ne j}\!\!\ls D_-
A^{ij}_+\rs\! D_-A^{ji}_++2\sum_i\!\lks A^{ii}_+\rks\!\!\lks\dd_k\dd_-
\lks A^{ii}_k\rks-ig\!\!\sum_{j', \,j'\ne i}\!
\ls A^{ij'}_kD_-A^{j'i}_k-\ls D_-A^{ij'}_k\rs A^{j'i}_k\rs\rks+\!\nop
+2\sum_{i,j,\, i\ne j}A^{ij}_+\ls\dd_kD_-A^{ji}_k-ig\sum_{j'}
\ls A^{jj'}_kD_-A^{j'i}_k-\ls D_-A^{jj'}_k\rs A^{j'i}_k\rs\rs-\frac{1}{2}
\sum_{i,j}F^{\phantom{kl}ij}_{kl}F^{klji}\Bigg\}+\nop
+\int d^2x\Bigg\{ 2L\sum_i(\dd_+v^i)^2-4L\sum_i\ls\dd_k A^{ii}_{k(0)}
\rs\dd_+v^i-2\sum_iA^{ii}_{+(0)}\bigg(2L\dd_k\dd_kv^i+\nor
+ig\ilm_{-L}^L
dx^-\sum_{j', \,j'\ne i}\ls A^{ij'}_kD_-A^{j'i}_k-\ls D_-A^{ij'}_k\rs A^{j'i}_k
\rs\bigg)\Bigg\}.
\nom}
Here we have ignored some unessential surface terms.

Variation of the Lagrangian in $[A_+^{ii}]$, $A_+^{ij}$ at $i\ne j$
leads to constraints, the solution of which can be written in the form
 \disn{2.43}{
[A_+^{ii}]=\dd_-^{-2}\lks
\dd_k\dd_-[A_k^{ii}]-ig\sum_{j',\, j'\ne i}\ls A_k^{ij'}D_-A_k^{j'i}-
(D_-A_k^{ij'})A_k^{j'i}\rs\rks,
\nom}
 \disn{2.44}{
A_+^{ij}\Bigr|_{i\ne j}=
D_-^{-2}\ls \dd_kD_-A_k^{ij}-ig\sum_{j'}
\ls A_k^{ij'}D_-A_k^{j'j}-(D_-A_k^{ij'})A_k^{j'j}\rs\rs.
\nom}
The operator $\dd_-^{-1}$ is completely defined, as before, by the condition
(\ref{2.24}) being well
defined on functions $[f(x)]$. The operator $D_-^{-1}$  after Fourier
transformation in $x^-$ is reduced to the multiplication by
$\ls i(k_- -g(v^i- v^j))\rs^{-1}$.
Therefore it has, in general, no singularities for
any $k_-=n(\pi /L)$ with integer n.

    Substituting the expressions (\ref{2.43}), (\ref{2.44})
into the Lagrangian (\ref{2.42}), we exclude
from it the quantities $[A_+^{ii}]$ and $A_+^{ij}$ at $i\ne j$. The variation
of the Lagrangian  in $A_{+(0)}^{ii}$ leads to the constraints
 \disn{2.45}{
Q^{ii}(x^+,x^\p)\equiv -2\ls
2L\dd_k\dd_k v^i+ig\ilm_{-L}^Ldx^-\sum_{j',\, j'\ne i}
\ls A_k^{ij'}D_-A_k^{j'i}-(D_-A_k^{ij'})A_k^{j'i}\rs\rs=0.
\nom}
They are first class constraints which can be posed on physical
state vectors after quantization. Therefore we can keep the term with this
constraint in the Lagrangian.

   Now we must put in the standard canonical form the terms of the Lagrangian
 \disn{2.46}{
\int d^2x\ilm_{-L}^L dx^-\left\{2\sum_i(\dd_-[A_k^{ii}])\dd_+[A_k^{ii}]+
2\sum_{i,j,\, i\ne j}(D_-A_k^{ij})\dd_+A_k^{ji}\right\}.
\nom}
It can be reached by going to Fourier transform
 \disn{2.47}{
[A_k^{ii}(x^-)]=\frac{1}{2\sqrt{2L}}\sum_{k_-=\pi/L}^\infty
k_-^{-1/2}\left\{ a_k^i(k_-)\exp(-ik_-x^-)+{a_k^i}^+(k_-)\exp(ik_-x^-)\right\},
\nom}
where we sum over $k_-=n\pi /L$, $n=1,2,\dots$, and
 \disn{2.48}{
A_k^{ij}(x^-)\Bigr|_{i\ne j}=
\frac{1}{2\sqrt{2L}}\left\{
\sum_{k_->g(v^i-v^j)}\ls k_--g(v^i-v^j)\rs^{-1/2}{a_k^{ij}}^+(k_-)
\exp(ik_-x^-)+\right.\nolr
+\left.\sum_{k_->g(v^j-v^i)}\ls k_--g(v^j-v^i)\rs^{-1/2}a_k^{ji}(k_-)
\exp(-ik_-x^-)\right\},
\nom}
where we sum over all $k_-=n\pi /L$ satisfying corresponding inequalities.
The expression (\ref{2.46}) takes the form
 \disn{2.49}{
(2i)^{-1}\int d^2x^\p\left\{
\sum_i\sum_{k_-=\pi/L}^\infty \ls a_k^i(k_-)\dd_+(a_k^i)^+(k_-)-
{a_k^i}^+(k_-)\dd_+a_k^i(k_-)\rs+\right.\nolr
\left.+\sum_{i,j,\, i\ne j}\sum_{k_->g(v^i-v^j)}
\ls a_k^{ij}(k_-)\dd_+{a_k^{ij}}^+(k_-)-
{a_k^{ij}}^+(k_-)\dd_+a_k^{ij}(k_-)\rs\right\}.
\nom}
  Further, in the Lagrangian (\ref{2.42}) there is a part
 \disn{2.50}{
L_v=\int d^2x^\p\left\{
2L\sum_i(\dd_+v^i)^2-4L\sum_i(\dd_kA_{k(0)}^{ii})\dd_+v^i\right\}.
\nom}
The "momentum" conjugated to $v^i$ is
 \disn{2.51}{
{\cal P}^i=\frac{\de L_v}{\de (\dd_+v^i)}=4L\ls \dd_+v^i-\dd_kA_{k(0)}^{ii}\rs,
\nom}
Hence,
 \disn{2.52}{
\dd_+v^i=\frac{1}{4L}{\cal P}^i+\dd_kA_{k(0)}^{ii}.
\nom}
The corresponding part of the Hamiltonian equals to
 \disn{2.53}{
H_v=\int d^2x^\p \sum_i\ls{\cal P}^i\dd_+v^i\rs-L_v=
\int d^2x^\p 2L\sum_i\ls \frac{{\cal P}^i}{4L}+\dd_kA_{k(0)}^{ii}\rs^2,
\nom}
and the corresponding part of canonical Lagrangian is
 \disn{2.54}{
L_v=\int d^2x^\p\sum_i\ls{\cal P}^i\dd_+ v^i\rs-H_v.
\nom}

Excluding from the Lagrangian (\ref{2.42}) the quantities
$[A_+^{ii}]$ and $A_+^{ij}$ (at $i\ne j$)
via the equations (\ref{2.47}), (\ref{2.48}),
replacing the
terms (\ref{2.46}) by the expression (\ref{2.49}) and
the part (\ref{2.50}) by the expression (\ref{2.54}), we obtain the result
 \disn{2.55}{
L=(2i)^{-1}\int d^2x^\p\left\{\sum_i\sum_{k_-=\pi/L}^\infty\ls
a_k^i(k_-)\dd_+{a_k^i}^+(k_-)-{a_k^i}^+(k_-)\dd_+a_k^i(k_-)\rs\right.+\nol
+\sum_{i,j,\, i\ne j}\sum_{k_->g(v^i-v^j)}\ls
a_k^{ij}(k_-)\dd_+{a_k^{ij}}^+(k_-)-{a_k^{ij}}^+(k_-)\dd_+a_k^{ij}(k_-)\rs+\nor
+\left.\sum_i {\cal P}^i\dd_+v^i+\sum_iA_{+(0)}^{ii}Q^{ii}\right\}-H,
\nom}
where $Q^{ii}$ are defined by (\ref{2.45}),
and the Hamiltonian $H=P_+$ is equal to
 \disn{2.56}{
H=\int d^2x\int\limits_{-L}^Ldx^-\left\{
\sum_i\ls \dd_-\lks A_+^{ii}\rks\rs^2+
\sum_{ij,\, i\ne j}\ls D_-A_+^{ij}\rs D_-A_+^{ji}-
\frac{1}{2}\sum_{i,j}F_{kl}^{\phantom{kl}ij}F^{klji}\right\}+\no
+2L\int d^2x^\p\sum_i\ls\frac{{\cal P}_i}{4L}+\dd_kA_{k(0)}^{ii}\rs^2.
\nom}

   It is implied that instead of the quantities $[A_+^{ii}]$ and $A_+^{ij}$
(at $i\ne j$) one uses the expressions (\ref{2.43}), (\ref{2.44})
and the $A_k^{ij}$ are
expressed in terms of $a_k^i$, ${a_k^i}^+$, $a_k^{ij}$, ${a_k^{ij}}^+$,
(at $i\ne j$) and of
$A_{k(0)}^{ii}$ with the help of equations (\ref{2.47}), (\ref{2.48}) and
 \disn{2.57}{
A_k^{ii}=\lks A_k^{ii}\rks +A_{k(0)}^{ii}.
\nom}

   It is seen from the formulae (\ref{2.55}) that ${a_k^i}^+$, $a_k^i$,
${a_k^{ij}}^+$, $a_k^{ij}$ play the role of
creation and annihilation operators. After quantization they satisfy the
following commutation relations (at $x^+=const$):
 \disn{2.58a}{
\lks a_k^i(k_-,x^\p),{a_l^j}^+(k'_-,x'^\p)\rks_-=\de^{ij}\de_{kl}
\de_{k_-,k'_-}\de^2(x^\p-x'^\p),
\nom}
 \disn{2.58b}{
\lks a_k^{ij}(k_-,x^\p),{a_l^{i'j'}}^+(k'_-,x'^\p)\rks_-=\de^{ii'}
\de^{jj'}\de_{kl}\de_{k_-,k'_-}\de^2(x^\p-x'^\p),\quad i\ne j,\;\; i'\ne j'.
\nom}
 Also we have
 \disn{2.58c}{
\lks {\cal P}^i(x^\p),v^j(x'^\p)\rks_-=-i\de^{ij}\de^2(x^\p-x'^\p).
\nom}
 Remaining commutators are equal to zero.

   The operator of the momentum $P_-$, defined by
 \disn{2.59}{
P_-=\int d^2x\int\limits_{-L}^Ldx^-T_{--},
\nom}
acts on physical states $|\Psi\ra$, satisfying the condition
 \disn{2.60}{
Q^{ii}(x^\p)|\Psi\ra=0,
\nom}
 equivalent to the  canonical operator
 \disn{2.61}{
P_-^{can}=\!\int\! d^2x\!\ls
\sum_i\sum_{k_-=\pi/L}^\infty\! k_-{a_k^i}^+(k_-)a_k^i(k_-)+
\sum_{i,j,\, i\ne j}\sum_{k_->g(v^i-v^j)}\! k_-
{a_k^{ij}}^+(k_-)a_k^{ij}(k_-)\rs,
\nom}
where the normal ordering was made.

   Physical vacuum $|\Omega\ra$ satisfies the relations
 \disn{2.62a}{
a_k^i(k_-,x^\p)|\Omega\ra=0,
\nom}
 \disn{2.62b}{
a_k^{ij}(k_-,x^\p)|\Omega\ra=0,\qquad i\ne j,
\nom}
and the condition (\ref{2.60}).

This scheme is connected with the following essential difficulty.
The zero modes   $A_{k(0)}^{ii}(x^\p\!)$
are present in the Lagrangian (\ref{2.55})
and in the Hamiltonian (\ref{2.56})  but the derivatives
$\dd_+  A_{k(0)}^{ii}$ are absent there. Therefore new constraints arise
 \disn{2.63}{
\frac{\de H}{\de A_{k(0)}^{ii}(x^\p)}=0.
\nom}
These constraints are of the 2nd class and  they must be solved with respect
to $A_{k(0)}^{ii}$  and then the $A_{k(0)}^{ii}$ have to be
excluded from the Hamiltonian. The constraints
(\ref{2.63})  are very complicated and  explicit resolution of them is practically
impossible. The application of Dirac brackets does not simplify this.

Due to this difficulty a practical calculation usually ignores all zero modes
from the beginning. It makes the approximation worse.   It is interesting that
in the framework of lattice regularization it is possible to overcome the
difficulties caused by the constraints (\ref{2.63}) \cite{heplat,tmf04}.
This question will be considered in sect.~5.

\section{Limiting transition from the theory in Lorentz\st
coordinates to the theory on the Light Front}

To  clarify  the  connection  between  the  theory  in    Lorentz
coordinates in Hamiltonian form and analogous theory on the LF
we perform the limiting transition from one to the other. Here
this transition
is considered in  the  fixed  frame  of  Lorentz  coordinates  by
introducing states that move at a speed close  to  the  speed  of
light in the direction of the $x^3$ axis. Constructing the matrix
elements of the Hamiltonian between such states and studying  the
limiting transition to the speed of light  (an
infinite  momentum),  we  can  derive  information   about    the
Hamiltonian in the light-like coordinates. This  information  also
takes into account the contribution from intermediate states with
finite momenta. Here,  we  illustrate  the  results  of  such  an
investigation using $(1+1)$-dimensional theory of scalar field with
the $\lambda \varphi^4$-interaction. Instead of $x^3$ we denote analogous
space coordinate by $x^1$. The generalization of the method to
$(3+1)$-dimensional Yukawa model is discussed briefly at the end of
this section.
The  limiting  transition   studied    here    is    accomplished
approximately by subjecting the momenta  $p_1$  to  an  auxiliary
cutoff that separates fast modes of the fields (with  high  $p_1$
values) from slow modes (with finite $p_1$ values).  This  cutoff
is parameterized in terms of the quantities  $\La$,  $\La_1$,  and
$\de$ and the limiting-transition parameter $\et$ ($\et >0$,
$\et \to 0$): we have $\et^{-1}\La_1\ge |p_1|\ge \et^{-1}\de$ for
the fast modes and $p_1\le \La$ for  the  slow  modes
($\La  \gg  \de$).  For   $\et    \to    0$,    the    inequality
$\et^{-1}\de>\La$ holds, so that the above momentum intervals are
separated.    The    field    modes    with    the        momenta
$\et^{-1}\de>|p_1|>\La$  are  discarded.  This    procedure    is
justified by the fact that the resulting Hamiltonian in the limit
$\et\to  0$  reproduces  the  canonical  LF  Hamiltonian
(without zero modes) when only the  fast  modes  are  taken  into
account and is consistent with conventional Feynman  perturbation
theory for $\de\to 0$. Therefore, even an  approximate  inclusion
of  the  other  (slow)  modes  may  provide  a  description    of
nonperturbative  effects,  such  as  vacuum   condensates.    The
effective LF Hamiltonian obtained  here  for  the  model
under consideration differs from the canonical  Hamiltonian  only
by the presence of the vacuum expectation  value  of  the  scalar
field and by an additional renormalization of the  mass  of  this
field. The renormalized  mass  involves  the  vacuum  expectation
value of the squared slow part of the field.
Masses of bound states can be found by solving Schrodinger equation
 \disn{3.1}{
P_+|\Psi\rangle=\frac{m^2}{2p_-}|\Psi\rangle,
\nom}
with obtained Hamiltonian $P_+$.

We start from the standard expression for the Hamiltonian of scalar
field $\varphi (x)$ in $(1+1)$-dimensional space-time in Lorentz
coordinates $x^{\mu}=(x^0,x^1)$, at $x^0=0$:
 \disn{3.2}{
H=:\int d^1x\ls \frac{1}{2}\Pi^2+\frac{1}{2}
(\dd_1\f)^2+\frac{m^2}{2}\f^2+\la\f^4\rs:,
\nom}
where $\Pi(x^1)$ are the variables that are canonically conjugate
to $\f(x^1)\equiv\f(x^0=0,x^1)$, and the symbol $:\quad :$ of the
normal ordering refers to the creation and annihilation operators
$a$ and $a^+$ that diagonalize the free part of  the  Hamiltonian
in the Fock space over the corresponding  vacuum  $|0\ra$.  These
operators are given by
 \disn{3.3}{
\f(x^1)=\frac{1}{\sqrt{4\pi}}\int dp_1(m^2+p^2_1)^{-1/4}
\lks a(p_1)\exp (-ip_1x^1)+h.c.\rks,
\nom}
 \disn{3.4}{
\Pi(x^1)=\frac{-i}{\sqrt{4\pi}}\int dp_1(m^2+p^2_1)^{-1/4}
\lks a(p_1)\exp (-ip_1x^1)-h.c.\rks,
\nom}
where $a(p_1)|0\ra=0$.

To  investigate  the  limiting  transition  to  the    LF
Hamiltonian (defined at $x^+= 0$), it is more convenient to go over
from the Hamiltonian (\ref{3.2}) to the operator
$H+P_1=\sqrt{2}P_+$, where the momentum $P_1$ has the form
 \disn{3.5}{
P_1=\int dp_1 a^+(p_1)a(p_1)p_1.
\nom}

Applying the above parametrization  of  high  momenta  in  terms of
$\et,\et\to 0$, to $p_1$ we can then consider  the  transition  to  an
infinitely  high  momentum  of  states  as  a  limit    of    the
corresponding Lorentz transformation with  parameter $\et$.  To  be
more specific, we have
$p_1\to (-\et\sqrt{2})^{-1}q_-$,
where  $q_-$  is  a  finite  momentum  in  the
light-like coordinates, and
 \disn{3.6}{
\lim_{\et\to 0}\ls(\et\sqrt{2})^{-1}\langle p'_1|(H+P_1)_{x^0=0}|p_1\ra\rs=
\langle q'_-|(P_+)_{x^+=0}|q_-\ra.
\nom}
It follows that the eigenvalues $E_+$ of  the  operator $(P_+)_{x^+=0}$  that
correspond to the momentum $q_-$ are
obtained as the corresponding limit of the eigenvalues  $E(\et)$  of
the operator $(H + P_1)_{x^0=0}$ at momentum $p_1$:
 \disn{3.7}{
E_+=\lim_{\et\to 0}(\et\sqrt{2})^{-1}E(\et).
\nom}
In the following, we consider this limiting transition as a  part
of the  eigenvalue  problem  for  the  operator  $H+P_1$,  using
perturbation theory in the parameter $\et$. Separating  the  Fourier
modes of the field into fast and slow ones, as  is  indicated
above,  and  neglecting  the  region  of  intermediate
momenta ($\et^{-1}\de\le |p_1|\le \et^{-1}\La_1$
is the region of  the  fast  modes,
and $|p_1|\le \La$ is the region
of  the  slow  modes),  we  can  substantially   simplify    this
perturbation theory. The $\et$ dependence of the field operators and
Hamiltonian can then be determined by making, in  the  region  of
fast momenta, the change of the variables as
 \disn{3.8}{
p_1=\et^{-1}k,\quad a(p_1)=\sqrt{\et}\tilde a(k),\quad
\de\le |k|\le \La_1,\quad [\tilde a(k),{\tilde a}^+(k')]=\de(k-k').
\nom}
The fast part $[\f(x^1)]_f$ of  the  operator
$\f(x^1)$ is estimated as
 \disn{3.9}{
[\f(x^1)]_f=\tilde \f(y)+O(\et^2),\quad y=\et^{-1}x^1,
\nom}
 \disn{3.10}{
\tilde\f(y)=(4\pi)^{-1/2}\int dk|k|^{-1/2}[\tilde a(k)\exp (-iky)+h.c.].
\nom}
We denote the slow part of the field $\f$ by $\breve\f$
($\f=[\f]_f+\breve\f$).
Substituting formulas (\ref{3.8})-(\ref{3.10}) into Hamiltonian (\ref{3.2}),
we obtain
 \disn{3.11}{
H+P_1=\et^{-1}\ls h_0+\et h_1+\et^2 h_2+\dots\rs,
\nom}
 \disn{3.12}{
h_0=2\int\limits_\de^{\La_1}dk\tilde a^+(k)\tilde a(k)k,
\nom}
 \disn{3.13}{
h_1=(H+P_1)_{\f=\breve\f,\Pi=\breve\Pi}\equiv (\breve H+\breve P_1),
\nom}
 \disn{3.14}{
h_2=\int\limits_{\de\le |k|\le \La_1}\hskip -3mm dk\ls\frac{m^2}{2|k|}\rs
\tilde a^+(k)\tilde a(k)+:\la\int dy\lks\tilde\f^4(y)+4\breve \f(0)\tilde\f^3(y)+
6\breve\f^2(0)\tilde\f^2(y)\rks:.
\nom}
Prior to performing integration with respect to  $y$,  we  formally
expanded the operators $\breve\f(x^1)=\breve\f(\et y)$
in Taylor  series  in  the
variable $\et y$ and estimated their orders in the parameter $\et$  at
fixed $y$. Such estimates  can  be justified  at  least  in  Feynman
perturbation theory. The
operator $(\breve H + \breve P_1)$ in (\ref{3.13})
is  defined  in  such  a  way  that  its
minimum eigenvalue is zero. Let us consider  perturbation  theory
in the parameter $\et$ for the equation
 \disn{3.15}{
(H+P_1)|f\ra=E|f\ra
\nom}
under the condition that the states $|f\ra$ have, as in formula (\ref{3.6}),
a negative value of $P_1$ proportional to $\et^{-1}$ for
$\et\to 0$ and also describe the states with a finite mass.
The expansions of the quantity $E$ and  the  vector  $|f\ra$  in  power
series in $\et$ can then be written as
 \disn{3.16}{
E=\et^{-1}\sum_{n=2}^\infty \et^nE_n,\qquad |f\ra=\sum_{n=0}^\infty \et^n|f_n\ra.
\nom}
We arrive at the system of equations
 \disn{3.17}{
h_0|f_0\ra=0,
\nom}
 \disn{3.18}{
h_0|f_1\ra+h_1|f_0\ra=0,
\nom}
 \disn{3.19}{
h_0|f_2\ra+h_1|f_1\ra+(h_2-E_2)|f_0\ra=0,\quad \dots.
\nom}
To describe  solutions  to
these equations, we use the basis  generated  by  the  fast-field
operators $\tilde a^+(k)$ over
the vacuum $|0\ra$ and the slow-field operators $\breve\f$ and $\breve\Pi$ over  the
vacuum $|v\ra$ that corresponds to the  Hamiltonian  $(\breve H +\breve P_1)$.
The vectors of this basis can be symbolically represented as
 \disn{3.20}{
\tilde a^+\dots\tilde a^+|0\ra
\breve\f\dots\breve\f\breve\Pi\dots\breve\Pi |v\ra.
\nom}
By virtue of (\ref{3.12}), the manifold of  solutions $|f_0\ra$  to  equation
(\ref{3.17}) is reduced to the set of vectors (\ref{3.20}), which do
not contain the operators $\tilde a^+(k)$ with $k\ge\de$.
Let  ${\cal P}_0$  be  the
projection operator onto this set. According to equation (\ref{3.18}), we
then have
 \disn{3.21}{
{\cal P}_0h_0|f_1\ra=(\breve H+\breve P_1)|f_0\ra=0.
\nom}
Equation (\ref{3.21}) requires that  the vectors
$|f_0\ra$ be the lowest
eigenstates of the operator $(\breve H+\breve P_1)$,
that is, linear combinations of
basis vectors (\ref{3.20}) including neither the
operators $\tilde a^+(k)$ with $k\ge\de$ nor the operators $\breve\f$ and
$\breve \Pi$. We denote
the projection operator on this set of vectors
(\ref{3.20}) by ${\cal P}'_0$. To determine the quantity $E_2$ which we are
interested in, it is sufficient to consider the ${\cal P}'_0$-projection of
equation (\ref{3.19}). Taking into account (\ref{3.17}), (\ref{3.20}),
and (\ref{3.21}), we  find
that $E_2$ appears as a solution to the eigenvalue problem
 \disn{3.22}{
{\cal P}'_0h_2|f_0\ra=E_2|f_0\ra.
\nom}
Thus, in accordance with (\ref{3.6})  and  (\ref{3.7}),  the
operator ${\cal P}'_0h_2{\cal P}'_0$ plays the  role  of  the
effective  LF
Hamiltonian $P_+^{eff}$. Substituting formula (\ref{3.14})  for
the  operator $h_2$
into the expression for $P_+^{eff}$, we take into account
that, between the projection operators ${\cal P}'_0$,  the  contribution  of
the field modes with positive momenta ($k\ge\de$)  vanishes  and  that
the products of the operators of the slow part of the  field  can
be replaced with their expectation values for the vacuum $|v\ra$.  In
addition,  we  note  that,  under  the  Lorentz    transformation
corresponding to the limiting transition $\et\to 0$ in formula  (\ref{3.6}),
the variable $y$ goes over into the light-like coordinate $y^-=-y/\sqrt{2}$,
the momenta $k$ go over into the light-like momenta $q_-=-\sqrt{2}k$, and  the
corresponding coordinate $y^+$ vanishes at $x^0=0$ (for finite values
of $y^-$). Going over to the operators
$A(q_-)=2^{-1/4}\tilde a(-k)$ for $k\le-\de$ \ \ ($q_-\ge\de\sqrt{2}$)
and  to  the  corresponding
field $\Phi(y^+=0,y^-)=\tilde\f_-(y)$, where $\tilde\f_-$
is the part  of  the  $\tilde\f$
containing only the modes with negative momenta  ($k\le-\de$),  we
obtain the effective Hamiltonian $P_+^{eff}$ in the form
 \disn{3.23}{
P_+^{eff}=:\int dy^-\left\{\frac{1}{2}\lks m^2+12\la\langle :\breve\f^2:\ra_v
\rks \Phi^2+4\la\langle \breve\f\ra_v\Phi^3+\la\Phi^4\right\}:,
\nom}
where $\langle\dots\ra_v$ is  the  expectation  value  for
the  vacuum $|v\ra$.
Expression  (\ref{3.23})  coincides  with   the    canonical    effective
Hamiltonian if, in the latter, we take into account the
shift of the field by the constant $\langle\breve\f\ra_v$
and the change in  the
mass squared by
$12\la\lks\langle :\breve\f^2:\ra_v-\langle\breve\f\ra^2_v\rks$.

Analogous results were obtained for Yukawa model in $(3+1)$-dimensional
space-time \cite{pred2} in regularization of Pauli-Villars type, introducing
a number of nonphysical fields with very large masses. The absence
of essential difference between the Hamiltonian obtained via limiting
transition and canonical LF Hamiltonian is connected with this choice
of regularization. Other regularizations can lead to more complicated
results.

This method of limiting transition can not be directly expanded to
gauge theories, because the approximations used for nongauge theories
are not justified.

\section{Comparison of Light Front perturbation theory\st
         with the theory in Lorentz coordinates}

As is already known, canonical quantization in LF, i.e.,
on the $x^+=const$ hypersurface, can result in a theory not  quite
equivalent to the Lorentz-invariant theory (i.e., to the standard
Feynman  formalism).  This  is  due,  first  of  all,  to  strong
singularities  at  zero  values  of  the  "light-like"   momentum
variables \hbox{$Q_-={1\over\sqrt{2}}(Q_0-  Q_3)$}.
To restore the  equivalence  with  a
Lorentz-covariant theory, one has to add unusual counter-terms to
the formal canonical Hamiltonian for the LF, $H=P_+$.  These
counter-terms can be found by comparing the  perturbation  theory
based  on  the  canonical  LF  formalism  with  Lorentz-covariant
perturbation theory \cite{tmf97}.
This is  done  in  the  present section.  The
LF  Hamiltonian  thus  obtained  can  then  be  used  in
nonperturbative  calculations.  It  is  possible,  however,  that
perturbation  theory  does  not  provide  all  of  the  necessary
additions  to  the  canonical  Hamiltonian,  as  some  of   these
additions can be nonperturbative. In spite  of  this,  it  seems
necessary  to  examine  this  problem  within  the  framework  of
perturbation theory first.

For practical purposes a stationary noncovariant
LF perturbation theory,
which is similar to  the  one  applied  in
nonrelativistic quantum mechanics, is widely used. It  was  found
\cite{bur1,har,lay}
that the "light-front" Dyson formalism allows  this  theory
to be transformed into an equivalent LF Feynman theory
(under an appropriate regularization). Then,  by  re-summing  the
integrands of the Feynman integrals, one can recast their form so
that they become the same as  in  the  Lorentz-covariant  theory.
(This is not the case for diagrams without external lines,  which
we do not  consider  here.)  Then,  the  difference  between  the
LF and Lorentz-covariant  approaches  that  persists  is
only due to the different regularizations and  different  methods
of calculating the Feynman integrals (which is important  because
of  the  possible  absence  of  their  absolute  convergence   in
pseudo-Euclidean space). In the present section, we concentrate  on
the analysis of this difference.

A  LF  theory  needs  not  only    the    standard    UV
regularization,  but  also  a  special  regularization  of   the
singularities $Q_-=0$. In  our  approach,  this  regularization
(by method (\ref{1.1}))
eliminates  the  creation  operators  $a^+(Q)$  and    annihilation
operators $a(Q)$ with $|Q_-^i|<\e$ from the Fourier expansion  of  the
field operators in the
field representation. As a result,  the  integration  w.r.t.  the
corresponding momentum $Q_-$ over the range $(-\infty,-\e)\cup(\e,\infty)$
is associated  with  each  line  before  removing  the  $\de$-functions.
Different propagators are regularized independently, which allows
the described re-arrangement of the perturbation  theory  series.
On the other hand, this regularization is convenient for  further
nonperturbative  numerical  calculations  with  the   LF
Hamiltonian, to which the necessary counter-terms are added  (the
"effective"  Hamiltonian).  We  require  that  this   Hamiltonian
generate a theory equivalent to the Lorentz-covariant theory when
the  regularization  is  removed.  Note  that   Lorentz-invariant
methods of regularization (e.g., Pauli-Villars regularization) are
far less convenient for numerical calculations and we shall  only
briefly mention them.

The specific properties  of  the  LF  Feynman  formalism
manifest themselves only in the integration over the variables
$Q_{\pm}={1\over\sqrt{2}}(Q_0\pm  Q_3)$,
 while integration over the transverse momenta $Q_{\p}\equiv \{Q_1, Q_2\}$
is  the  same  in  the  LF  and  the    Lorentz
coordinates (though it might be nontrivial  because  it  requires
regularization and renormalization). Therefore, we concentrate on
a comparison of diagrams for fixed transverse momenta  (which  is
equivalent to a two-dimensional problem).

In this section we propose a method that allows one to find
the difference (in the limit $\e\to 0$)  between  any  LF
Feynman integral and the corresponding Lorentz-covariant integral
without having  to  calculate  them  completely.  Based  on  this
method, a procedure is elaborated for constructing  an  effective
LF Hamiltonian  correct to any order of  perturbation theory.  This
procedure can be applied to nongauge field theories as  well
as to Abelian and non-Abelian gauge theories in the gauge $A_-=0$
with the gauge vector field  propagator  chosen  according  to  the
Mandelstam-Leibbrandt prescription \cite{man,lei}.
The question of whether  the
additions to the Hamiltonian, that arise under that procedure, can  be
combined into a finite number of counter-terms must be considered
separately in each particular case.

We will consider the application of this  formalism
to Yukawa  model, to QCD in four-dimensional space-time
and to QED in two-dimensional space-time.

\subsection{Reduction of Light Front and Lorentz-covariant
Feynman\st integrals to a form convenient for comparison}
\label{integ}

Let us examine an arbitrary one particle irreducible
Feynman diagram. We fix all external momenta and all transverse
momenta of integration, and integrate only over
$Q_+$ and $Q_-$:
 \disn{4.1}{
F=\lim_{\ka\to 0}\int{{\prod_i d^2Q^i \quad  f(Q^i,p^k)}
\over  {\prod_i(2Q_+^iQ_-^i-M^2_i+i\ka)}}.
\nom}
We assume that all vertices are polynomial and that the propagator has the form
 \disn{4.1.2}{
{{z(Q)}\over {Q^2-m^2+i\ka}},\qquad{\rm or}\qquad {{z(Q)\; Q_+}
\over{(Q^2-m^2+i\ka)(2Q_+Q_-+i\ka)}},
\nom}
where $z(Q)$ is a polynomial.
A propagator of the second type in (\ref{4.1.2}) arises in gauge theories
in the gauge  $A_-=0$
if the Mandelstam-Leibbrandt formalism \cite{man,lei}
with the vector field propagator
 \disn{nnn5}{
\frac{1}{Q^2+i\ka}\ls g_{\mu\nu}-
\frac{Q_\mu\de_\nu^++Q_\nu\de_\mu^+}{2Q_+Q_-+i\ka}2Q_+\rs,
\nom}
is used.
In equation (\ref{4.1}) either $M^2_i=m^2_i+{Q^{i}_{\p}}^2\ne 0$, where  $m_i$
is the particle mass, or $M^2_i=0$.

The function  $f$
involves the numerators of all propagators and all vertices
with the necessary
$\de$-functions,
that include the external momenta  $p^k$  (the same expression without the
$\de$-functions is a polynomial, which we denote by $\tilde  f$).
We assume for the diagram  $F$  and for all of its subdiagrams that
the conditions
 \disn{4.1.1}{
\om_{\pa}<0, \qquad \om_+<0,
\nom}
hold, where $\om_+$ is the index of divergence w.r.t.
 $Q_+$ at $Q_-^i\ne 0\  \forall  i$,  and  $\om_{\pa}$
is the index of divergence in $Q_+$ and $Q_-$ (simultaneously);
 \hbox{$Q_{\pm}={1\over\sqrt{2}}(Q_0\pm  Q_3)$}.
The diagrams that do not meet these conditions should
be examined separately for each particular theory
(their number is usually finite). We
seek the difference
between the value of integral
(\ref{4.1})
obtained by the Lorentz-covariant calculation and its value calculated in
LF coordinates (LF calculation).

In the LF calculation, one introduces and
then removes the LF cutoff
$|Q_-|\ge\e>0$:
 \disn{nnn6}{
F_{\rm lf}=\lim_{\e\to 0}\lim_{\ka\to 0}
\int\limits_{V_{\scriptstyle \e}}\prod_i dQ_-^i \int
\prod_i dQ_+^i
{{f(Q^i,p^k)}\over{\prod_i (2Q_+^iQ_-^i-M^2_i+i\ka)}},
\nom}
where
 $V_{\e}=\prod_i\ls\ls -\infty,-\e\rs \cup\ls \e,\infty\rs \rs $.
Here (and in the diagram configurations to be defined below) we take
the limit w.r.t. $\e$,
but, generally speaking, this limit may not exist. In this case, we assume
that we do not
take the limit, but take the sum of all nonpositive power terms of the
Laurent series in $\e$ at the zero point.
If conditions
(\ref{4.1.1}) are satisfied, Statement 2 from Appendix I can be used.
This results in the equality
 \disn{4.5}{
F_{\rm lf}=\lim_{\e\to 0}\lim_{\ka\to 0}\int\prod_k dq_+^k
\int\limits_{V_{\scriptstyle \e}\cap B_L} \prod_k dq_-^k
{{\tilde f(Q^i,p^s)}
\over{\prod_i (2Q_+^iQ_-^i-M^2_i+i\ka)}}.
\nom}
From here on, the momenta of the lines $Q^i$
are assumed to be expressed in terms of the loop momenta
$q^k$,  $B_L$  is a sphere of a radius  $L$ in the $q_-^k$-space,  and
$L$ depends on the external momenta. Now, using Statement 2 from
Appendix I, we obtain
 \disn{4.7}{
F_{\rm lf}=\lim_{\e\to 0}\lim_{\ka\to 0}
\lim_{\bet\to 0}\lim_{\g\to 0}
\int\prod_k dq_+^k \int\limits_{V_{\scriptstyle \e}}
\prod_k dq_-^k
{{\tilde f(Q^i,p^s) \; e^{-\g\sum_i {Q_+^i}^2-\bet \sum_i{Q_-^i}^2}
}\over{\prod_i (2Q_+^iQ_-^i-M^2_i+i\ka)}}.
\nom}
To reduce the covariant Feynman integral to a form similar to
(\ref{4.5}), we introduce a quantity $\hat F$:
 \disn{4.8}{
\hat F=\lim_{\ka\to  0}\lim_{\bet\to 0}\lim_{\g\to 0}
\int  \prod_k  d^2q^k  {{\tilde  f(Q^i,p^s)  \;
e^{-\g\sum_i    {Q_+^i}^2-\bet
\sum_i{Q_-^i}^2}}\over{\prod_i (2Q_+^iQ_-^i-M^2_i+i\ka)}}.
\nom}
Let us prove that this quantity coincides with the result of the
Lorentz-covariant calculation $F_{\rm  cov}$. To this end, we introduce the
\hbox{$\al$-representation}  in the Minkowski space of the propagator
 \disn{4.9}{
{{z(Q^i)}\over{2Q_+^iQ_-^i-M^2_i+i\ka}}
=-iz\ls -i{{\dd}\over{\dd y_i}}\rs \int\limits_0^{\infty}
e^{i\al_i(2Q_+^iQ_-^i-M^2_i+i\ka)+i(Q^i_+y_i^++Q^i_-y_i^-)}
d\al_i \Bigr|_{y_i=0}.
\nom}
Then we substitute (\ref{4.9}) into (\ref{4.8}). Due to the exponentials
that cut off $q_+^k$, $q_-^k$ and $\al^i$ the integral over these
variables is absolutely convergent. Therefore, one can interchange
the integrations over $q_+^k$,
$q_-^k$ and $\al^i$. As
a result, we obtain the equality
 \disn{4.18}{
\hat F=\lim_{\ka\to 0}\lim_{\bet\to 0}\lim_{\g\to 0}
\int\limits_0^\infty \prod_n d\al_i \;\hat \f(\al_i,p^s,\g,\bet)
\; e^{-\ka\sum_i\al_i},
\nom}
where
 \disn{4.19}{
\hat \f(\al_i,p^s,\g,\bet)=(-i)^n
\tilde f\ls -i{{\dd}\over{\dd y_i}}\rs \times \nolr
\times \int \prod_k d^2q^k \;
e^{\sum_i \lks i\al_i(2Q_+^iQ_-^i-M^2_i)+
i(Q^i_+y_i^++Q^i_-y_i^-)-\g{Q_+^i}^2-\bet{Q_-^i}^2\rks }
\Bigr|_{y_i=0}.
\nom}
For the Lorentz-covariant calculation in the \hbox{$\al$-representation}
satisfying conditions (\ref{4.1.1}), there is a known
expression \cite{bo}
 \disn{4.17}{
F_{\rm cov}=\lim_{\ka\to 0}
\int\limits_0^\infty \prod_n d\al_i \;\f_{\rm cov}(\al_i,p^s)
\; e^{-\ka\sum_i\al_i},
\nom}
where
 \disn{4.16}{
\f_{\rm cov}(\al_i,p^s)=(-i)^n \tilde f\ls -i{{\dd}\over
{\dd y_i}}\rs \times\nolr
\times \lim_{\g,\bet\to 0} \int \prod_k d^2q^k
\; e^{\sum_i \lks i\al_i(2Q_+^iQ_-^i-M^2_i)+
i(Q^i_+y_i^++Q^i_-y_i^-)-\g{Q_+^i}^2-\bet{Q_-^i}^2\rks }
\Bigr|_{y_i=0}.
\nom}
In Appendix 2, it is shown that in (\ref{4.18}) the limits in
$\g$ and $\bet$ can be interchanged, in turn, with the integration
over
$\{\al_i\}$, and then with $\tilde f\ls -i{{\dd}\over{\dd  y_i}}\rs $.
After that, a comparison of relations (\ref{4.18}), (\ref{4.19}) and (\ref{4.17}),
(\ref{4.16}),  clearly shows that $\hat F=F_{\rm   cov}$.
Considering  (\ref{4.8})  and using Statement 1 from Appendix 1,
we obtain the equality
 \disn{4.20}{
F_{\rm cov}=\lim_{\ka\to  0}\int\prod_k  dq_+^k
\int\limits_{B_L}
\prod_k    dq_-^k    {{\tilde    f(Q^i,p^s)}
\over{\prod_i (2Q_+^iQ_-^i-M^2_i+i\ka)}}.
\nom}
Expression (\ref{4.20}) differs from  (\ref{4.5})
only by the range of the integration over $q_-^k$.

\subsection{Reduction of the difference between the Light Front
and\st Lorentz-covariant Feynman integrals
to a sum of configurations}
\label{polos}

Let us introduce a partition for each line,
 \disn{4.22}{
\ls  \int\limits_{-\infty}^{-\e}
dQ_- + \int\limits_{\e}^{\infty} dQ_- \rs =
\lks \int dQ_-+(-1)\int\limits_{-\e}^{\e} dQ_-\rks .
\nom}
We call a line with integration w.r.t. the momentum  $Q_-^i$  in  the
range $(-\e,\e)$ (before removing the $\de$-functions) a type-1 line, a
line with integration in the range $(-\infty,-\e)\cup(\e,\infty)$
a  type-2
line, and a line with integration over the whole range
$(-\infty,\infty)$
a full line. In the  diagrams,  they  are  denoted  as  shown  in
Figs.~la, b, and c, respectively.
 \begin{figure}[ht]
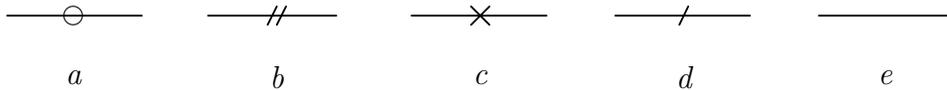

\hskip 13mm\input fig1.pic
\caption{Notation for different types of lines  in  the  diagrams:
"a" is a type-1 line, "b" is a type-2 line, "c" is a  full  line,
"d" is an $\e$-line, and "e" is a $\Pi$-line.}
\end{figure}

Let us substitute partition (\ref{4.22}) into expression (\ref{4.5})
for $F_{\rm  lf}$ and
open the brackets. Among the resulting terms,  there  is  $F_{\rm cov}$
(expression  (\ref{4.20})).  We  call  the  remaining   terms    "diagram
configurations" and denote them by $F_j$. Then  we  arrive  at  the
relation  $F_{\rm lf}-F_{\rm cov}=\sum\limits_jF_j$, where
 \disn{4.24}{
F_j=\lim_{\e\to 0}\lim_{\ka\to 0}\int\prod_k dq_+^k
\int\limits_{V^j_{\scriptstyle \e}\cap B_L} \prod_k dq_-^k
{{\tilde f(Q^i,p^s)}
\over{\prod_i (2Q_+^iQ_-^i-M^2_i+i\ka)}},
\nom}
and $V^j_{\e}$ is the region corresponding to  the  arrangement  of  full
lines and type-1 lines in the given configuration.

Note that before taking the limit in $\e$,
equations (\ref{4.20}) and  (\ref{4.24})
can be used successfully: first, they are applied to a  subdiagram  and,
then, are substituted into the formula for  the  entire  diagram.
This is admissible because, after the deformation of the contours
described in the proof  of  Statement  1  from  Appendix  1,  the
integral over the loop momenta $\{q_+^k\}$ of the  subdiagram  converges
(after integration over the variables $\{q_-^k\}$  of  this  subdiagram)
absolutely and uniformly  with  respect  to  the  remaining  loop
momenta $\{q_-^{k'}\}$. Therefore, one can interchange the  integrals  over
$\{q_+^k\}$ and $\{q_-^{k'}\}$.

Thus,   the    difference    between    the    LF    and
Lorentz-covariant calculations of the diagram is given by the sum
of all of its configurations. A configuration of a diagram is the
same diagram, but where each line is labeled as a full or  type-1
line, provided that at least one type-1 line exists.

\subsection{Behavior of the configuration as  $\e\to 0$}
\label{epsil}

We  assume  that  all
external momenta $p^s$ are fixed for the diagram in question and
 \disn{4.D0}{
p_-^s\ne 0, \quad \sum_{s'}p_-^{s'}\ne 0,
\nom}
where the summation is taken over any subset  of  external  momenta;
all of these momenta are assumed to be directed inward.

Let us consider an arbitrary configuration.  We  apply  the  term
"$\e$-line" to all type-1 lines  and  those  full  lines  for  which
integration over $Q_-$ actually does not  expand  outside  the  domain
$(-r\e,r\e)$, where  $r$ is a finite number  (below,  we  explain  when
these lines appear). The remaining full lines are called $\Pi$-lines.
In the diagrams, the $\e$-lines and $\Pi$-lines are denoted as shown  in
Figs.~1d and e, respectively. Note that the diagram can be  drawn
with  lines  "a"  and  "c"  from  Fig.~1  (this   defines    the
configuration unambiguously), or with lines "d" and "e" (then the
configuration is not uniquely defined).

If among the lines arriving at the vertex only one is  full
and the others are type-1 lines, this full line is an  $\e$-line  by
virtue of the momentum conservation at the vertex. The  remaining
full lines form a subdiagram (probably unconnected). By virtue of
conditions (\ref{4.D0}), there is a connected part to which  all  of  the
external lines are attached. All of the  external  lines  of  the
remaining  connected  parts  are   $\e$-lines.  Consequently,   using
Statement~1 from Appendix~1, we can see that integration over the
internal momenta of these connected parts can be carried out in a
domain of order $\e$ in size, i.e., all of their internal lines  are
$\e$-lines. Thus, an arbitrary configuration can be drawn as in Fig.~2
and integral (\ref{4.24}), with the corresponding  integration  domain,
is associated with it.
 \begin{figure}[ht]
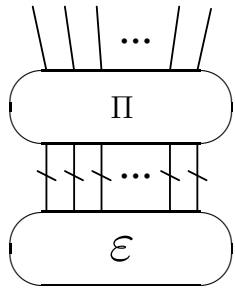

\hskip 11mm\input fig2.pic
\caption{
Form of an arbitrary configuration: $\Pi$ is  the  connected
subdiagram consisting of $\Pi$-lines, {\protect\LARGE $\e$}
is the subdiagram  consisting
of $\e$-lines and, probably, containing no vertices.}
\end{figure}

Let us investigate the behavior of the configuration  as  $\e\to 0$.
From here on, it is convenient to represent the propagator as
 \disn{4.D1}{
{{\tilde z(Q)}\over {Q^2-m^2+i\ka}},\quad{\rm where}\quad
 \tilde z(Q)=z(Q)\; \; {\rm or}\; \; \tilde z(Q)={{z(Q)}\over
{2Q_-+i\ka/Q_+}}.
\nom}
rather than as (\ref{4.1.2}). Then, in (\ref{4.1}),
$M^2_i=m^2_i+{Q^{i}_{\p}}^2\ne 0$  and  the
function $\tilde f$ is no longer a polynomial. If  the  numerator  of  the
integrand consists  of  several  terms,  we  consider  each  term
separately (except when  the  terms  arise  from  expressing  the
propagator momentum $Q_-^i$ in terms of loop and external momenta).

We denote the loop momenta of subdiagram $\Pi$ in Fig.~2 by  $q^l$ and
the others  by  $k^m$.  We  make  following  change  of  integration
variables in (\ref{4.24}):
 \disn{4.D10}{
k_-^m \to \e \; k_-^m.
\nom}
Then,  the  integration  over  $k_-^m$  goes  within  finite    limits
independent of $\e$. We denote the power of $\e$ in the  common  factor
by $\ta$ (it stems from the volume elements  and  the  numerators
when the transformation (\ref{4.D10}) is made). The contribution to $\ta$ from the expression
$1/(2Q_-+i\ka/Q_+)$
(equation (\ref{4.D1})), which is related to the $\e$-line, is equal  to  -1.  We
divide the domain of integration over $k_+^m$ and $q_+^l$ into sectors such
that the momenta of all full lines $Q_+^i$ have the same  sign  within
one sector.

In Statement~1 of Appendix~1, it is shown that for  each  sector,
the contours of integration over $q_-^l$ and $k_-^m$ can be bent in such a
way that absolute convergence in $q_+^l$,  $k_+^m$,
$q_-^l$  and $k_-^m$ takes place.
Since, in this case, the momenta $Q_-^i$  of  $\Pi$-lines  are  separated
from  zero  by  an $\e$-independent  constant,  the   corresponding
$\Pi$-line-related propagators and factors from the vertices  can  be
expanded in a series in $\e$. This expansion commutes with integration.

It is also clear that the denominators of the  propagators  allow
the following estimates under an infinite increase in $|Q_+|$:
$$\displaylines{
\hfill
\left|{1\over{2Q_+ Q_--M^2+i\ka}}\right|\le \cases{
{\displaystyle {1\over{c \; |Q_+|}}} & for $\Pi$-lines,
\refstepcounter{equation} \label{4.D11} \hskip 30mm \hfill (\theequation)\cr
{\displaystyle {1\over{\tilde c\; \e \; |Q_+|}}} & for $\e$-lines,
\refstepcounter{equation} \label{4.D12} \hfill (\theequation)\cr}
}$$
Here $c$ and $\tilde c$ are $\e$-independent constants.
Note  that  for  fixed
finite $Q_+$, the estimated expressions are bounded as $\e\to  0$. After
transformation (\ref{4.D10}) and release of the factor ${1\over{\e}}$
(in  accordance
with  what  was  said  about  the  contribution  to $\ta$),    the
$\e$-line-related expression from (\ref{4.D1}) becomes
 \disn{nnn7}{
\left| {1\over{2Q_-+i\ka/Q_+}}\right| \to
\left| {1\over{2Q_-+i\ka/(Q_+\e)}}\right| \le
{1\over{2|Q_-|}},
\nom}
where a $Q_+$-independent quantity was used for the  estimate  (this
quantity is meaningful and does not  depend  on  $\e$  because  the
value of $Q_-$ is separated from zero by an $\e$-independent constant).

We integrate first over $q_+^l$, $k_+^m$ within one sector  and  then  over
$q_-^l$,  $k_-^m$  (the latter integral converges uniformly in  $\e$).
Let  us
examine the  convergence  of  the  integral  over  $q_+^l$,  $k_+^m$   with
canceled denominators of the $\e$-lines (which is equivalent to
estimating the expressions (\ref{4.D12}) by a constant). If it converges, then
the initial integral  is  obviously  independent  of  $\e$  and  the
contribution  from  this  sector  to   the    configuration    is
proportional to $\e^\ta$.

Let us show that if it diverges with a degree of  divergence  $\al$,
the contribution to the initial integral is proportional to $\e^{\ta-\al}$
up to logarithmic corrections. To this end, we divide the  domain
of integration over $q_+^l$,
$k_+^m$ into  two  regions:  $U_1$,  which  lies
inside a sphere of radius $\La/\e$ ($\La$ is fixed), and $U_2$,  which  lies
outside this sphere (recall that in our reasoning, we  deal  with
each sector separately). Now we estimate (\ref{4.D11})
(like  (\ref{4.D12}))  in
terms of ${\displaystyle{1\over{\hat  c\e|Q_+|}}}$
(which is admissible) and change  the  integration
variables as follows:
 \disn{4.D13}{
q_+^l\to {1\over{\e}}\; q_+^l,\quad k_+^m\to {1\over{\e}}\; k_+^m.
\nom}
After $\e$ is factored out of the numerator and the volume  element,
the integrand becomes  independent  of  $\e$.  Thus,  the  integral
converges.

One can choose such $\La$ (independent of  $\e$) that  the  contribution
from the  domain  $U_2$ is  smaller  in  absolute  value  than  the
contribution from the domain $U_1$. Consequently, the whole integral
can he estimated
via the integral over the  finite  domain  $U_1$.  Now  we  make  an
inverse replacement in (\ref{4.D13}) and estimate (\ref{4.D12})
by a  constant  (as
above). Since the size of the integration domain is $\La/\e$  and  the
degree of divergence is $\al$, the integral behaves as  $\e^{-\al}$  (up  to
logarithmic corrections), q.e.d. This reasoning is valid for each
sector and, thus, for the configuration as a whole. Obviously,
 \disn{4.D13.1}{
\al=\max\limits_{r}\al_r,
\nom}
where $\al_r$ is the subdiagram divergence index and the  maximum  is
taken over all subdiagrams $D_r$ (including unconnected  subdiagrams
for which $\al_r$ is the sum  of  the  divergence  indices  of  their
connected parts). In the case under consideration, $\al_r=\om_+^r+\n^r$,
where $\n^r$ is the number of internal $\e$-lines in the subdiagram
$D_r$.
The  quantities  $\om_{\pm}^r$  are  the  UV  divergence  indices  of   the
subdiagram  $D_r$ w.r.t. $Q_{\pm}$.

Above, we introduced a quantity $\ta$, which is equal to the power of
$\e$ that stems from the  numerators  and  volume  elements  of  the
entire configuration. We can write $\ta=\om_-^r-\m^r+\n^r+\eta^r$,  where
$\m^r$ is the index of the UV  divergence  in  $Q_-$   of  a smaller subdiagram
(probably, a tree subdiagram or a nonconnected one) consisting of
$\Pi$-lines entering $D_r$. The term $\eta^r$
is the power of $\e$ in the  common
factor, which, during transformation (\ref{4.D10}), stems from the  volume
elements and numerators of the lines that did not enter $D_r$. (It is
implied that the integration momenta are chosen in the  same  way
as when calculating the divergence indices of $D_r$.)  Then,  up  to
logarithmic corrections, we have
 \disn{4.13.4}{
F_j\sim \e^{\s},\quad \s=\min_r(\ta,\om_-^r-\om_+^r-\m^r+\eta^r).
\nom}
Consequently, for $\e\to 0$, the configuration is equal to zero if
$\s>0$. Relation (\ref{4.13.4})
allows  all  essential  configurations  to  be
distinguished.

\subsection{Correction procedure and analysis of counter-terms}
\label{ispra}

We want to build a corrected LF Hamiltonian
$H_{\rm  lf}^{\rm  cor}$ with  the
cutoff $|Q_-^i|>\e$, which would generate  Green's  functions  that
coincide in the limit $\e  \to  0$  with  covariant  Green's  functions
within the perturbation theory. We begin with a  usual  canonical
Hamiltonian in the LF coordinates  $H_{\rm lf}$  with  the  cutoff
$|Q_-^i|>\e$. We imply that the integrands of the  Feynman  diagrams
derived from  this  LF  Hamiltonian  coincide  with  the
covariant integrands after some  resummation  \cite{bur1,har,lay}.
However,  a
difference may arise due to the  various  methods  of  doing  the
integration, e.g., due to different auxiliary regularizations. As
shown in Sec.~\ref{polos}, this difference (in the limit
$\e  \to  0$) is equal to
the sum of all properly arranged configurations of  the  diagram.
One should  add  such  correcting  counter-terms  to  $H_{\rm lf}$,  which
generates  additional  "counter-term"  diagrams,  that  reproduce
nonzero (after taking limit w.r.t. $\e$) configurations  of  all  of
the diagrams. Were we able  to  do  this,  we  would  obtain  the
desired $H_{\rm lf}^{\rm cor}$.
In fact, we can only show how to  seek  the  $H_{\rm lf}^{\rm cor}$  that
generates the Green's functions  coinciding  with  the  covariant
ones everywhere except the null  set  in  the  external  momentum
space (defined by condition (\ref{4.D0})). However, this  restriction  is
not essential because this possible difference  does  not  affect
the physical results.

Our  correction  procedure  is  similar  to  the  renormalization
procedure. We assume that the perturbation  theory  parameter  is
the number of loops. We carry out the correction by steps: first,
we find the counterterms to the  Hamiltonian  that  generate  all
nonzero configurations of the diagrams up to the given order and,
then, pass to the next order. We take into account that this step
involves  the  counter-term  diagrams  that  arose    from    the
counter-terms added to the Hamiltonian for lower orders. Thus, at
each  step,  we  introduce  new  correcting  counter-terms   that
generate the difference remaining in this order. Let us show  how
to successfully look for the correcting counter-terms.

We call a configuration nonzero if it does not vanish as $\e\to 0$.
We  call  a  nonzero  configuration  "primary"  if $\Pi$  is  a   tree
subdiagram in it (see Fig.~2). Note that for this  configuration,
breaking any  $\Pi$-line results in a violation of  conditions  (\ref{4.D0});
then, the resulting diagram is not a configuration. We  say  that
the configuration is changed if all of the $\Pi$-lines in the related
integral (\ref{4.24})  are expanded in series  in  $\e$  (see  the  reasoning
above equation (\ref{4.D11}) in Sec.~\ref{epsil}) and only those terms that
do not vanish
in the limit $\e\to 0$ after the
integration are retained. As  mentioned  above,  developing  this
series and integration are interchangeable operations.  Thus,  in
the limit $\e  \to  0$,  the  changed  and  unchanged  configurations
coincide. Therefore,  we  always  require  that  the  Hamiltonian
counter-terms generate changed configurations, as this simplifies
the form of the counter-terms.  Using  additional  terms  in  the
Hamiltonian, one can generate only counter-term  diagrams,  which
are equal to zero for  external  momenta  meeting  the  condition
$|p_-^s|<\e$, because with the cutoff used,
the external lines of the diagrams  do  not  carry  momenta  with
$|p_-^s|<\e$. We bear this in mind in what follows.

We seek counter-terms by the induction method. It is clear  that,
in  the  first  order  in  the  number  of  loops,  all   nonzero
configurations  are  primary.  We  add  the  counter-terms   that
generate them to the Hamiltonian. Now, we  examine  an  arbitrary
order of perturbation theory. We assume that in lower orders, all
nonzero  configurations  that  can   be    derived    from    the
counter-terms, accounting for the  above  comment,  have  already
been generated by the Hamiltonian.

Let us proceed to  the  order  in  question.  First,  we  examine
nonzero configurations with only one loop momentum $k$ and a number
of momenta $q$ (see the notation above  equation (\ref{4.D10})).  We  break  the
configuration lines one by one without touching the  other  lines
(so that the ends of the broken lines become external lines). The
line break may result in a structure that is not a  configuration
(if conditions (\ref{4.D0}) are violated); a line break may  also  result
in a zero configuration or in a  nonzero  configuration.  If  the
first case is realized for each broken line, then  the  initial
configuration  is  primary  and  it  must  be  generated  by  the
counter-terms  of  the  Hamiltonian   in    the    order    under
consideration. If breaking of each line results in either  the  first
or the second case, we call the initial configuration real and it
must be also generated in this order.

Assume that breaking a line results in the third case. This means
that the resulting configuration stems from counter-terms in the
lower orders. Then, after restoration of the broken  line  (i.e.,
after  the  appropriate  integration),  it  turns  out  that  the
counter-terms of the lower  orders  have  generated  the  initial
configuration (we take into account  the  comment  on  successive
application of equation (\ref{4.24});  see  the  end  of
Sec.~\ref{polos})  with  the
following distinctions: (i) the broken line (and, probably,  some
others, if a nonsimply connected diagram  arises  after  breaking
the line) is  not  a  $\Pi$-line  but  a  type-2  line,  due  to  the
conditions $|p_-^s|>\e$; (ii) if, after restoration  of  the  broken
line, the behavior at small $\e$ becomes worse (i.e., $\s$ decreased),
then fewer terms than are necessary for the initial configuration
were considered in the above-mentioned series  in  $\e$.  We  expand
these arising type-2 lines by formula  (\ref{4.22})  and  obtain  a  term
where all of these lines are replaced by $\Pi$-lines or  other  terms
where some (or all) of these lines have become type-1  lines.  In
the latter case, one of the momenta $q$ becomes the momentum $k$.  We
call these  terms  "repeated  parts  of  the  configuration"  and
analyze them together  with  the  configurations  that  have  two
momenta  $k$.  In  the  former  case,  we  obtain    the    initial
configuration up to distinction (ii). We add  a  counter-term  to
the  Hamiltonian  that  compensates  this  distinction    (the
counter-term diagrams generated by it are called the compensating
diagrams).

If there is only one line for which the third case  is  realized,
it turns out that, in the given order, it is  not  necessary  to
generate the initial configuration by the  counter-terms,  except
for the compensating addition  and  the  repeated  part  that  is
considered at the next step. If there are several lines for which
the  third  case  is  realized,  the  initial  configuration   is
generated in lower orders more than once.  For  compensation,  it
should be generated (with the corresponding numerical coefficient
and the opposite sign) by the Hamiltonian  counter-terms  in  the
given order. We call this configuration a secondary one. Next, we
proceed to examine configurations with two momenta $k$ and so on up
to  configurations  with  all  momenta  $k$,  which  are   primary
configurations.

Thus, the configurations  to  be  generated  by  the  Hamiltonian
counter-terms can be  primary  (not  only  the  initial  primary
configurations but also the repeated parts  analogous  to  them,
called primary-like),
real, compensating, and  secondary.  If  the  theory  does  not
produce either the loop consisting only of lines with $Q_+$  in  the
numerator (accounting for contributions from the vertices)  or  a
line with ${Q_+}^n$ in the numerator for $n>1$, then real configurations
are absent because a line without $Q_+$ in the numerator can  always
be broken without  increasing $\s$  (see  equation (\ref{4.13.4})).  It  is  not
difficult to demonstrate that if  each appearing  primary, real,  and
compensating configuration has only two external line,
then there are  no  secondary configurations  at  all.

The  dependence    of    the    primary
configuration on external momenta becomes trivial if its degree of
divergence $\al$ is positive, the maximum in formula (\ref{4.D13.1})
is  reached
on the diagram itself, and $\s=0$. Then, only the first  term  is
taken into account in the above-mentioned series. Thus,  not  all
of the $\Pi$-line-related propagators and vertex  factors  depend  on
$k_-^m$ and they can be pulled out of the sign of the integral w.r.t.
$\{k_-^m\}$ in (\ref{4.24}). We then obtain
 \disn{4.D14}{
F_j^{\rm prim}=\lim_{\e\to  0}\lim_{\ka\to 0}\int\prod_m dk_+^m
{{\tilde f'(k^m,p^s)}\over{\prod_i (2Q_+^iQ_-^i-M^2_i+i\ka)}} \times\nolr
\times \int\limits_{V_{\scriptstyle \e}}  \prod_m   dk_-^m
{{\tilde f''(k^m)} \over{\prod_k (2Q_+^kQ_-^k-M^2_k+i\ka)}},
\nom}
where $V_{\e}$ is a domain  of  order  $\e$  in  size.  Let  us  carry  out
transformations (\ref{4.D10}) and (\ref{4.D13}).
For the denominator of the $\Pi$-line,
we obtain
 \disn{nnn8}{
{1\over{2({1\over{\e}}\sum k_+ +\sum p_+)(\sum p_-)-M^2+i\ka}}\to
{{\e}\over{2(\sum k_+)(\sum p_-)}}.
\nom}
Here we neglect terms of order $\e$ in the denominator  because  the
singularity at $k_+^m=0$ is integrable under the  given  conditions
for $\al$ and everything can be calculated in zero order in $\e$ at $\s=0$.
Thus, the dependence on external  momenta  can  be  completely
collected into an easily obtained common factor.

\subsection{Application to the Yukawa model}
\label{jukav}

The Yukawa model involves diagrams that do not satisfy  condition
(\ref{4.1.1}). These are displayed in Figs.~3a~and~b. We have
$\om_{\pa}=0$  for
diagram "a" and $\om_+=0$ for diagram "b".
 \begin{figure}[ht]
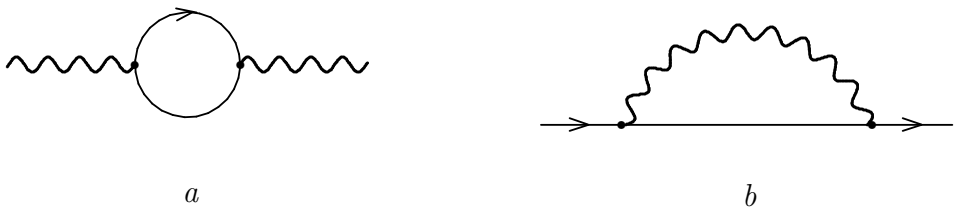

\hskip 13mm\input fig3.pic
\caption{Yukawa model diagrams that do not meet condition
(\protect\ref{4.1.1}).}
\end{figure}

Nevertheless, these  diagrams  can  be  easily  included  in  the
general scheme of reasoning. To this end, one should subtract the
divergent part, independent of external momenta, in the integrand
of the logarithmically divergent (in two-dimensional space,  with
fixed internal transverse momenta)  diagram  "a".  We  obtain  an
expression with $\om_{\pa}<0$ (i.e., which converges  in  two-dimensional
space) and $\om_+=0$, as  in  diagram  "b".  This  means  that  the
integral over $q_+$ converges only in the  sense  of  the  principal
value (and it is this value of the integral that should be  taken
in the LF  coordinates  to  ensure  agreement  with  the
stationary noncovariant perturbation theory). This value  can  be
obtained by distinguishing the $q_+$-even part of the integrand.

Two approaches are possible. One is to introduce  an  appropriate
regularization in transverse momenta  and  to  imply  integration
over them; then, it is convenient to distinguish the part that is
even in four-dimensional momenta $q$.  The  other  is  to  keep  all
transverse  momenta  fixed;  then,  the  part  that  is  even  in
longitudinal momenta $q_{\pa}$ can be released. For the  Yukawa  theory,
we use the first approach. For the transverse regularization,  we
use a "smearing" of vertices, which  is  equivalent  to  dividing
each propagator by $1+{Q_{\p}^i}^2/{\La_{\p}}^2$.
In  four-dimensional  space,
diagram  "a"  diverges  quadratically.  Under  introduction   and
subsequent  removal  of  the  transverse   regularization,    the
divergent  part,  which  was  previously  subtracted  from   this
diagram, acquires the form $C_1+C_2\> p_{\p}^2$.

After separating the even part of the regularized expression,  we
fix all of the transverse momenta again. Then it turns  out  that
diagrams "a" and "b" in Fig.~3 meet conditions (\ref{4.1.1})  and  one  can
show that after all of the operations mentioned, the exponent  $\s$
(see (\ref{4.13.4})) does not decrease for  any  of  their  configurations.
Hence, they can be included in the  general  scheme  without  any
additional corrections.

Let  us  first  analyze  the  primary  configurations  (see   the
definition in Sec.~\ref{ispra}). In the numerators, $k_-$  appears  only  in
the zero or one power and there are no loops where the numerators
of all of the lines contain $k_-$. Consequently, one always has
$\ta>0$,  $\m^r\le  0$, and $\eta^r\ge 0$ (see the definitions in
Sec.~\ref{epsil}).  Analyzing the
properties of the expression $\om_-^r-\om_+^r$ for the Yukawa model diagrams,
we conclude from (\ref{4.13.4}) that $\s\ge  0$ always holds. The general  form
of the nonzero primary configurations with $\s=0$ is depicted in
Fig.~4. Note that they are all configurations with two external line.
 \begin{figure}[ht]
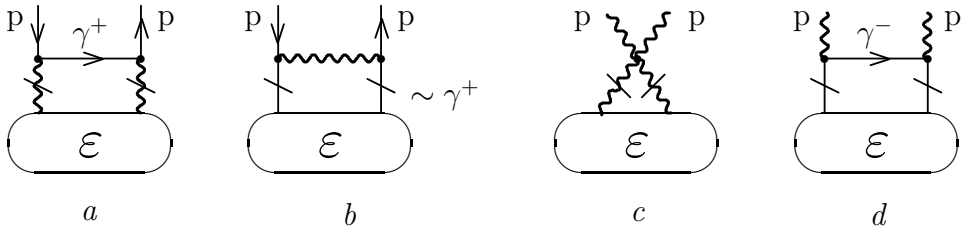

\hskip 12mm\input fig4.pic
\caption{
Nonzero configurations in the Yukawa model: $p$ is the external
momentum, and $\g^+$ or $\g^-$
symbols on the line indicate that the corresponding term is taken
in the numerator of the propagator.
In configuration "b", the part that is proportional to $\g^+$ is taken.}
\end{figure}

Further,  it  is  clear  that  there  are   no    nonzero    real
configurations (see the comment at the end of Sec.~\ref{ispra}), and it can
be shown by induction that there are no nonzero  compensating  or
secondary configurations either (the  definitions  are  given  in
Sec.~\ref{ispra} also). Thus, only primary or  primary-like  configurations
can be nonzero and all of them have the form shown in Fig.~4.  It
can be shown that their degree of divergence $\al$ is  positive  and
the maximum in formula (\ref{4.D13.1}) is reached for  the  diagram  itself.
Thus, the reasoning above and below formula (\ref{4.D14}) applies to them.
Then, denoting the configurations  displayed  in  Figs.~4a-d  by
\hbox{$D_{a}$~--~$D_{d}$}, we
arrive at the equalities
\hbox{$D_a={{\g^+}\over{p_-}}C_a$},
\hbox{$D_b={{\g^+}\over{p_-}}C_b$},             \hbox{$D_c=C_c$}
and \hbox{$D_d=C_d$},
where the expressions \hbox{$C_a$ -- $C_d$} depend only on the  masses  and
transverse momenta, but not on the external longitudinal momenta,
and have a finite limit as $\e\to 0$.

Now we assume that $D_{a}$  --  $D_{d}$
are not single  configurations  but  are
the sums  of  all  configurations  of  the  same  form  and  that
integration over the internal transverse momenta has already been
carried  out,  (with  the  above-described  regularization).   In
four-dimensional space, the diagrams $D_{a}$ and $D_{b}$ diverge  linearly
while $D_{c}$ and $D_{d}$ diverge quadratically. Therefore, because of  the
transverse regularization, the coefficients  $C_c$  and  $C_d$  in  the
limit of removing this regularization take the form $C_1+C_2\>  p_{\p}^2$,
where $C_1$ and $C_2$ do not depend on the external momenta (neither do
$C_a$, $C_b$)). Thus, to generate all  nonzero  configurations  by  the
LF Hamiltonian, only the expression
 \disn{4.U1}{
H_c=\tilde C_1\; \f^2+\tilde C_2\> p_{\p}^2\; \f^2+
\tilde C_3\; \bar \psi\; {{\g^+}\over{p_-}}\; \psi,
\nom}
should be added, where $\f$ and $\psi$ are the boson and fermion fields,
respectively, and $\tilde C_i$, are the constant coefficients.

Comparing  (\ref{4.U1}) with   the    initial    canonical LF
Hamiltonian, one can easily see that the found counter-terms  are
reduced to a renormalization of various terms of the  Hamiltonian
(in particular the boson mass squared  and  the  fermion  mass
squared without changing the term linear in fermion mass).  The explicit
Lorentz invariance is absent but  results are Lorentz invariant because
the counterterms  compensate the violation  of
Lorentz invariance inherent to  chosen LF formalism.

Note that the  second approach,  mentioned  at
the beginning of this section,  can give the  same  results.
The  only  difference  is  that  in  two-dimensional  space,  the
contributions from the configurations displayed in Fig.~3  would
additionally depend on external transverse momenta. However, this
dependence disappears after integration over internal  transverse
momenta with  the  introduction  and  subsequent  removal  of  an
appropriate regularization.

In the Pauli Villars regularization, it is easy  to  verify  that
the expression $\om_-^r-\om_+^r-\m^r+\eta^r$ from (\ref{4.13.4})
increases. This is
because the number of terms in the numerators of  the  propagator
increases. Then, the  contribution  from  the  $\e$-lines  does  not
change, while the $\Pi$-lines belonging to $D_r$ make  zero  contribution
to $\om_-^r-\om_+^r$ and $\eta^r$, but $-1$ contribution to $\m^r$.
Since $\ta>0$, this
regularization makes it possible to meet
the condition $\s>0$ for the  configurations  that  were  nonzero
(one additional boson field and one additional fermion field  are
enough).  Then  it  turns  out  that  the  canonical  LF
Hamiltonian need not be corrected at all.

Obtained results agree with the conclusions of
the work  \cite{bur1}, where
a comparison of   LF  and  Lorentz-covariant
methods was made for self-energy  diagrams  in  all  orders  of
perturbation  theory  and  for other  diagrams  in   lowest    orders.

\subsection{Application to QCD(3+1)}
\label{kalib}

Applying the LF Hamiltonian approach to gauge theories
under regularization (\ref{1.1}),
where zero modes of fields are thrown out, one
has to use the gauge $A_-=0$ (see, for example, sect.~2.3).
To carry out successfully renormalization procedure for this scheme
it is necessary  to take gauge field propagator
in accordance with Mandelstam-Leibbrandt
prescription \cite{man,lei} (which allows to perform
Euclidean continuation, see \cite{bas1,bas2,skark}).
Such a propagator has the form
 \disn{4n1}{
\frac{-i\de^{ab}}{Q^2+i\ka}
\ls g_{\m\n}-\frac{Q_\mu\de_\nu^++Q_\nu\de_\mu^+}{2Q_+Q_-+i\ka}2Q_+\rs=
\frac{-i\de^{ab}}{Q^2+i\ka}
\ls g_{\m\n}-\frac{Q_\m n_\n+Q_\n n_\m}{2(Qn^*)(Qn)+i\ka}2(Qn^*)\rs,
\nom}
where $n_+=1$, $n_-=n_\p=0$, $n^*_-=1$, $n^*_+=n^*_\p=0$.

The formalism described in sect.~4.1-4.4 was such that it could be
applied to a theory with the propagator (\ref{4n1})
(at fixed transverse momenta
$Q_{\p}\ne 0$). It turns out that there  are  nonzero  configurations  with
arbitrarily large numbers of external lines. An example of such a
configuration is given in Fig.~5.
 \begin{figure}[ht]
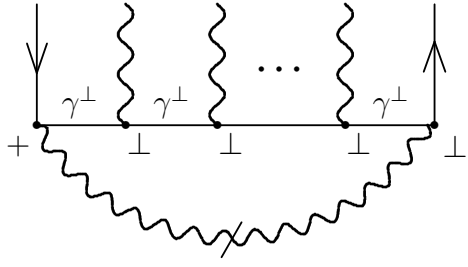

\hskip 10mm\input fig5.pic
\caption{
Nonzero configuration with an arbitrarily large number of external
lines in a gauge theory.
The symbols $\g^{\p}$ on the lines and the symbols $+$ or $\p$
the vertices indicate that the corresponding
terms $\g^+$ or $\g^\p$ are taken in the numerators of
propagators and in the vertex factors.}
\end{figure}

Indeed, using formula (\ref{4.13.4}), we can see that for the configuration
in Fig.~5, $\ta=0$ and, thus, $\s\le  0$,  i.e.,  this  is  a  nonzero
configuration.  It  is  also  clear  that  introduction  of   the
Pauli-Villars  regularization  does  not  improve  the  situation
because it does not affect $\ta$.

The difficulty  is  that
the distortion of the pole in (\ref{4n1})
due to LF cutoff $|Q_-|\ge\e >0$ does
not disappear in the limit $\e\to 0$, and infinite number of new
counterterms are required to compensate this distortion
\cite{tmf97}. The simplest way to avoid this difficulty is to add
small mass-like parameter $\mu^2$ in the denominator:
 \disn{4n16}{
 \frac{1}{2Q_+Q_-+i\ka}\longrightarrow  \frac{1}{2Q_+Q_--\mu^2+i\ka},
 \nom}
and take the limit $\e\to 0$ before $\mu\to 0$). To describe this
modification with local Lagrangian we need to introduce ghost
fields $A'_{\mu}$ in addition to conventional $A_{\mu}$. We write
the free part of pure gluon Lagrangian as follows (using higher
derivatives and the parameter $\La$ for UV regularization):
 \disn{4n17}{
 L_0=-\frac{1}{4}\Biggl( f^{a,\mu\n} \ls 1+\frac{\dd^2}{\La^2}\rs
 f_{\mu\n}^a-f'^{a,\mu\n}
 \ls 1+\frac{\dd^2}{\La^2}+\frac{2\dd_+\dd_-}{\mu^2}\Biggr)
 f'^a_{\mu\n}\rs,
 \nom}
where $f^a_{\mu\n}=\dd_\mu A^a_{\n}-\dd_\n A^a_{\mu}$,
$f'^a_{\mu\n}=\dd_\mu A'^a_{\n}-\dd_\n A'^a_{\mu}$ and
$A^a_-=A'^a_-=0$. Interaction terms depend only on summary field
$\bar A^a_{\mu}=A^a_{\mu}+A'^a_{\mu}$.

At fixed $\mu$ and $\La$  we get a theory with broken gauge
invariance but with preserved global  $SU(3)$-invariance. We put
into the Lagrangian all necessary interaction terms (but with
unknown  coefficients) including those that are needed
for UV renormalization:
 \disn{4n18}{
L=L_0+c_0\dd_\mu\bar A^a_\n\dd^\mu\bar A^{a,\n}+
c_{01}\dd_\mu\bar A^a_\n n^\m n^{*\al}\dd^\al\bar A^{a,\n}+
c_1\dd_\mu\bar A^{a,\mu}\,\dd_\n\bar A^{a,\n}+\nol+
c_{11}n^\m n^{*\al}\dd_\mu\bar A^{a,\al}\,\dd_\n\bar A^{a,\n}+
c_{12}n^\m n^{*\al}\dd_\mu\bar A^{a,\al}
n^\n n^{*\bet}\dd_\n\bar A^a_\bet+
c_2\bar A^a_\mu\bar A^{a,\mu}+\no+
c_3 f^{abc}\bar A^a_\mu\bar A^b_\n \dd^\mu\bar A^{c,\n}+
c_{31} f^{abc}\bar A^a_\mu\bar A^b_\n
n^\al n^{*\m}\dd^\al\bar A^{c,\n}+\no+
\bar A^a_\mu\bar A^b_\n\bar A^c_\g\bar A^d_\de
\bigl(c_4 f^{abe}f^{cde} g^{\mu\g}g^{\n \de}+
\de^{ab}\de^{cd}\ls c_5 g^{\mu\g}g^{\n \de}+
c_6 g^{\mu\n}g^{\g\de}\bigr)\rs.
 \nom}
These terms are local and can be taken in
Lorentz covariant form due to the restoring of this symmetry in
the $\e\to 0$, $\mu \to 0$ limit.
However in constructing of these terms both vectors
 $n^\m$ and $n^{*\n}$, included in the definition of the propagator
(\ref{4n1}), can participate.

For this theory one can apply   the formalism described
 in sect.~4.1-4.4  to compare (at $\e\to 0$)
the LF perturbation theory
and that taken in Lorentz coordinates within the same
regularization scheme.  We find that the difference between
mentioned perturbation theories can be compensated by changing of
the value of coefficient $c_2$ before the term of gluon mass form
$\bar A^a_\mu\bar A^{\mu,a}$ in naive LF Hamiltonian of this
theory. After that we can analyse further our regularized theory
in Lorentz coordinates and even make Euclidean continuation.

It is possible to prove by induction to all orders of perturbation theory
that in the limit $\mu\to 0$, $\Lambda\to\infty$ our theory can be
made finite and coinciding with the usual renormalized
(dimensionally regularized) theory in Light Cone gauge \cite{bas1,bas2}
(for all Green functions). This was done in the paper \cite{tmf99},
but there the terms of the Lagrangian, including  the vector
 $n^{*\m}$, were missed. Right expression for
LF Hamiltonian should correctly take into account the contribution of
all terms, written in the (\ref{4n18}).

The values of the unknown coefficients $c_i$ before all counterterms in
(\ref{4n18}) must be chosen so that the
Green functions in each order coincided (after removing the
regularization) with those obtained in conventional dimensionally
regularized formulation and therefore satisfied Ward identities.
Besides, we need to correlate the limits $\mu\to 0$ and
$\La\to\infty$ to avoid infrared divergencies at $\mu\to 0$. It is
sufficient to take $\mu=\mu(\La)$ and to require that $\mu\La\to
0$ and $(\log\mu)/\La\to 0$.

Our resulting LF Hamiltonian for pure $SU(3)$ gluon fields
contains 11 unknown coefficients, including coefficient before
gluon mass term that takes also into account the difference
between LF and Lorentz coordinate formulations of our regularized
theory. The generalization of our scheme for full QCD with
fermions is described in \cite{tmf99}. In this case
there are 20 unknown coefficients in the LF Hamiltonian. We
hope that it is possible to find an analog of Ward identities
relating the coefficients $c_i$ at fixed $\La$. This problem seems
very important for our approach.

\subsection{Application to QED(1+1)}
The procedure, described in previous section, allows
to construct LF Hamiltonian for QCD in four-dimensional
space-time, but such LF Hamiltonian contains
many additional fields and unknown coefficients.
This complicates  calculations
with this Hamiltonian. Moreover, because only the perturbation theory
with respect to the coupling constant was analyzed, there could
remain purely nonperturbative effects that are not taken into
consideration.
It is therefore useful  to consider an example
of "nonperturbative"
(with respect to usual coupling constant)
 construction of LF Hamiltonian for gauge theory,
which is possible for two-dimensional quantum electrodynamics (QED(1+1)),
i.e. for "massive" Schwinger model.

The QED(1+1), defined originally by the Lagrangian
 \disn{4d1}{
 L=-\frac{1}{4}F_{\mu\n}F^{\mu\n}+\bar\Psi(i\g^mD_\mu-M)\Psi,
 \nom}
can be transformed to its bosonized form \cite{colm,naus}, described by
scalar field Lagrangian
 \disn{4d2}{
 L=\frac{1}{2}\ls\dd_\mu\Phi\dd^\mu\Phi-m^2\Phi^2\rs+
 \frac{Mme^C}{2\pi}\cos(\te+\sqrt{4\pi}\Phi),
 \nom}
where $m=e/\sqrt{\pi}$ \ is the Schwinger boson mass (the $e$ is
original coupling), $C=0.577216\dots$ is the Euler constant,
and the $\te$ is the
"$\te$"-vacuum parameter, which takes into account the nontriviality of
QED(1+1) quantum vacuum due to instantons.
Here the fermion mass $M$ plays the
role of the coupling in bosonized theory so that perturbation
theory in this coupling corresponds to chiral perturbation theory in
QED(1+1). The nonpolynomial form of scalar field interaction leads
in  perturbation theory to infinite sums of diagrams in each
finite order. It can be proved \cite{shw2,tmf03} that some partial sums
of these infinite sums are UV divergent in the 2nd order, whereas
for full (Lorentz-covariant) Green functions these divergencies
cancel (remaining only the divergent vacuum diagrams).  Therefore physical
quantities are UV finite in this theory. Only at intermediate
steps of our analysis we need some UV regularization.

We compare LF and Lorentz-covariant perturbation theories for such
bosonized model using an effective resummation of
perturbation series in coordinate representation for Feynman
diagrams \cite{shw2,tmf02} and also using the formalism
described in sect.~4.1-4.4.
The results of this comparison can be formulated as
follows.

The difference between  considered perturbation theories can be
eliminated in the limit of removing regularizations if we use instead
of the  naive LF Hamiltonian
 \disn{4d3}{
 H=\int dx^-\ls\frac{1}{8\pi}\,m^2:\f^2:
 -\frac{\g}{2}\,e^{i\te}:e^{i\f}:-
 \frac{\g}{2}\,e^{-i\te}:e^{-i\f}:\rs,\no
 \g=\frac{Mme^C}{2\pi},\quad
 \f=\sqrt{4\pi}\,\Phi,\quad
 |p_-|\ge\e>0,
 \nom}
the "corrected" LF Hamiltonian:
 \disn{4d4}{
 H=\int dx^-\ls\frac{1}{8\pi}\,m^2:\f^2:
 -B:e^{i\f}:-B^*:e^{-i\f}:\rs-\nol
 -2\pi e^{-2C}\frac{|B|^2}{m^2}
 \int dx^-dy^- \ls
 :e^{i\f(x^-)}e^{-i\f(y^-)}:-1\rs\te(|x^--y^-|-\al)
 \frac{v(\e(x^--y^-))}{|x^--y^-|}.
 \nom}
Here the terms, linear in $B$ and $B^*$  (new coupling constants),
are of the same form as in naive Hamiltonian; only  the term,
containing the $|B|^2$, is of new form (nonlocal in $x^-$). The
$\al$ is the UV regularization parameter, and the $v(z)$ is some
arbitrary continuous function rapidly decreasing at the infinity and
going to unity as $z\to 0$. The coupling $B$ can be perturbatively
written as a series in $\g$:
 \disn{4d5}{
 B=\frac{\g}{2}e^{i\te}+\sum_{n=2}^{\infty}\g^nB_n.
 \nom}
On the other side, it is related to the sum of all connected
"generalized tadpole" diagrams (i.e. diagrams with
external lines attached to only one vertex), which is described by
the "condensate" parameter
$A=\frac{\g}{2}\langle\Om|:e^{i(\f+\te)}:|\Om\ra$ of the
Lorentz-covariant formulation (the $|\Om\ra$ is the physical vacuum
state in this formulation):
 \disn{4d5.1}{
 B+|B|^2w=A,
 \nom}
 \disn{4d6}{
 w=\frac{2\pi e^{-2C}}{m^2}\int dx^-\frac{\te(|x^-|-\e\al)}{|x^-|}v(x^-).
 \nom}
The equation (\ref{4d5.1}) can be solved with respect to the $B$:
 \disn{4d6.1}{
 B=-\frac{1}{2w}+\sqrt{\frac{1}{4w^2}+\frac{A'}{w}-A''^2}+iA'',
 \nom}
where $A=A'+iA''$, and the sign before the root respects the
perturbation theory. Within the perturbation theory in $\g$ one
cannot remove UV regularization (i.~e. to put $\al\to 0$ and therefore
$w\to\infty$) in this expression due to UV divergencies of the
coefficients $B_n$. However, taking into account the validity of
the equation (\ref{4d5.1}) to all orders in $\g$, we can consider it
beyond the perturbation theory. Then we use the estimation for the
$A$ at $\al\to 0$ \cite{shw2}:
 \disn{4d6.2}{
 A=\frac{\g^2}{4}w+const
 \nom}
and get for the $B$ in $\al\to 0$ limit UV finite result:
 \disn{4d7}{
B={\rm sign}(\cos\te)\sqrt{\frac{\g^2}{4}-A''^2}+iA''=
\frac{\g}{2}e^{i\hat\te},
 \nom}
so that all information about the condensate is contained in the phase
factor $e^{i\hat\te}$:
 \disn{4d8}{
 \sin\hat\te=2\frac{{\rm Im}A}{\g}=\langle\Om|:\sin(\f+\te):|\Om\ra.
 \nom}

Then we can make a transformation, inverse to the
bosonization, but on the LF. Actually we need the expression only
for one independent component $\psi_+ $ of the bispinor field
$\ls\psi_+\atop\psi_-\rs$ due to the LF constraint, permitting to
write the $\psi_-$ in terms of $\psi_+$. One can use the exact
expression for the $\psi_+$ in terms of the $\f$ obtained in the
theory on the interval $|x^-|\le L$ with periodic boundary
conditions \cite{naus,shw2}. We need only to modify our corrected
bosonized theory by using discretized LF momentum $p_-$ instead of
continuous one and hence replacing the cutoff parameter $\e$ by
$\pi/L$. The necessary formulae for the $\psi_+$ has the following
form \cite{shw2,naus} (we choose here antiperiodic boundary
conditions for the fermion fields):
 \disn{4d9}{
 \psi_+(x)=\frac{1}{\sqrt{2L}}e^{-i \om}e^{-i\frac{\pi}{L}x^-  Q}
 e^{i\frac{\pi}{2L}x^-}:e^{-i\f(x)}:.
 \nom}
The operator $\om$ and the  charge operator $Q$ are canonically
conjugated so that the $\psi_+$ defined by the equation (\ref{4d9})  has
proper commutation relation with the charge. On the other side the
operator $e^{i\om}$  shifts Fourier modes $\psi_n$ of the field $\psi_+$
\cite{shw2,naus}:
 \disn{4d10}{
 e^{i\om}\psi_n e^{-i\om}=\psi_{n+1}.
 \nom}
If we separate the modes related with creation and annihilation
operators on the LF putting
 \disn{4d11}{
 \psi_+(x)=\frac{1}{\sqrt{2L}}\Biggl(
 \sum_{n\ge 1}b_ne^{-i\frac{\pi}{L}(n-\frac{1}{2})x^-}+
 \sum_{n\ge 0}d_n^+ e^{i\frac{\pi}{L}(n+\frac{1}{2})x^-}\Biggr),\quad
 b_n|0\ra=d_n|0\ra=0,
 \nom}
we can define the operator $e^{i\om}$  uniquely by specifying its
action on the LF vacuum $|0\ra$ as follows:
 \disn{4d12}{
 e^{i \om}|0\ra=b^+_1 |0\ra,\qquad
 e^{-i \om}|0\ra=d^+_0 |0\ra.
 \nom}
In such sense this operator is similar to a fermion.

We can now rewrite our corrected boson LF Hamiltonian in terms of
$\ps_+$ and  $e^{i\om}$. The result is remarkably simple:
 \disn{4d13}{
 H=\int\limits_{-L}^Ldx^-
 \biggl(\frac{e^2}{2}\ls \dd_-^{-1}
 [\psi_+^+ \psi_+]\rs^2-\frac{iM^2}{2}
 \psi_+^+\dd_-^{-1} \psi_+
 -\ls\frac{Me\:e^C}{4\pi^{3/2}} e^{-i\hat\te}\:e^{i \om}d_0^+
 +h.c.\rs\biggr).
 \nom}
This fermionic LF Hamiltonian differs from canonical one (in
corresponding DLCQ scheme) only by last term, depending on zero
modes and vacuum condensate parameter $\hat\te$ which can be
related to chiral condensate by transforming the variables in
the equation (\ref{4d8}):
 \disn{4d14}{
 \sin\hat\te=-\frac{2\pi^{3/2}}{e\;e^C}
 \langle\Om|:\bar\Psi\g^5\Psi:|\Om\ra.
 \nom}

Let us remark that the presents of linear in $M$ term in
LF Hamiltonian (\ref{4d13}) can be considered as a results
of a modification of the LF constraint, connecting
the  $\psi_-$ with the $\psi_+$. An analogous
modification of this constraint was got in the paper
\cite{mac} where the method of
exact operator solution of massless Schwinger model was applied.

The constructed LF Hamiltonian (\ref{4d13}) was applied to
nonperturbative numerical calculation of mass spectrum
of QED(1+1) \cite{rasch}.
The results of this calculation were compared with those of
lattice calculations in Lorentz coordinates
\cite{ham1,ham2}.

The calculations were carried out in wide domain of values of fermion mass $M$
for all values of the parameter $\hat\te$, which is a function
of the $M/e$ and vacuum parameter $\te$.
We do not know exactly this function, but know that it must be zero
at $\te=0$ and be equal to $\pi$ at  $\te=\pi$.

For  $\te = 0$  the obtained spectrum is  bounded from  below
at any values $M$, and is in good agreement with the results of the paper
\cite{ham1} for two bound states of lowest mass.

For the value $\te=\pi$, at which phase transition can take place,
the obtained spectrum for the lowest bound state agrees well with
the results of the paper \cite{ham2}  for sufficiently small  $M$;
at greater  $M$ we start to see the disagreement with that paper, and then
at $M$, greater some critical value, the spectrum
becomes unbounded from below.
This critical value approximately coincides with the point of phase
transition found in \cite{ham2}. It can be supposed that
at $M$ greater than the point of phase transition the calculations
with our LF Hamiltonian, constructed via the analysis of perturbation theory
in $M$, become incorrect.

Our calculations show that LF Hamiltonian (\ref{4d13}) can give
good results only in the limited domain of the parameters: from
perturbative domain to some values
which define the limits of applicability of our LF Hamiltonian.
In the domain where our Hamiltonian becomes incorrect  we need to take into
account nonperturbative (in M/e) effects.  Such an investigation can be useful
for finding a LF Hamiltonian approach to realistic gauge theories.

The LF Hamiltonian (\ref{4d13}) includes the operator
$\om$, which has no simple expression in terms of field operators.
It is defined only by its properties (\ref{4d10}),(\ref{4d12}).
Due to this fact the expression for the Hamiltonian depends essentially
on the form of the regularization, i.e. $|x^-|\le L$ and
antiperiodic boundary conditions in $x^-$ for the
field $\psi$.
Now we have found a possibility to rewrite the expression (\ref{4d13})
in such a way that it contains only fermion field operators,
and describes in the limit of removing the regularizations the same theory
as the Hamiltonian (\ref{4d13}).
This new expression has at $\hat\te=\te=0$ the following form:
 \disn{new1}{
H=\int\limits_{-L}^Ldx^-\ls\frac{e^2}{2}\ls \dd_-^{-1}
[\psi^+\psi]\rs^2
+\frac{eMe^C}{4\pi^{3/2}}\ls d_0^+d_0+b_1^+b_1\rs-
\frac{iM^2}{2}\psi^+\dd_-^{-1}\psi\rs.
\nom}
Preliminary calculations of the mass spectrum, produced by this Hamiltonian,
show that in the limit of removing the regularization, $L\to\infty$, results
indeed coinside, with a good accuracy, with those for the bound state mass
spectrum found here for the Hamiltonian (\ref{4d13}).
Work on studing of the Hamiltonian (\ref{new1}) spectrum and also
on the constructing the analogous expression for the Hamiltonian
at $\hat\te\ne0$  will be continued in future.

\section{Transverse lattice regularization of Gauge Theories\st
on the  Light Front}

The introduction of space-time lattice for gauge-invariant regularization
of nonabelian gauge theories is well known \cite{wils}.
Gauge invariant regularization in continuous space-time is also known
\cite{halp} but it seems not suitable for the LF quantization.
For the LF formulation only
the lattice in transverse coordinates $x^1, x^2$ is used.
In this formulation
it is convenient to define variables so as to have the action
polynomial in these variables \cite{barpir1,barpir2,heplat,tmf04}.
Such a regularization is not Lorentz
invariant, and one can only hope that Lorentz invariance
can be restored in continuous space limit.
Nevertheless many attempts to apply LF Hamiltonian formulation
with periodic boundary conditions in $x^-$,
combined with transverse space
lattice, are undertaken (for "color dielectric" type models
\cite{dalley1,dalley4,pirner1,mack}).
In all of these works zero modes of fields are thrown out,
so that, in fact, gauge invariance is violated.

We consider canonical  LF formulation of gauge theories,
regularized in gauge-invariant way. To achieve this goal we
introduce transverse space
lattice, discretize the momentum $p_-$ according to the prescription
(b) (see introduction, equation (\ref{1.2}))
with all zero modes of fields included and apply the so called
"finite mode" ultraviolet
regularization in $p_-$. The last means a cutoff in eigenvalues of
covariant derivative operator $D_-$  in the expansion of lattice field
variables in eigen functions of this operator. These field
variables are  lattice modification of transverse components of usual gauge
fields. They are described by complex matrices, defined on lattice links.
Only such variables
admit mentioned above "finite mode" regularization (for fermion fields
analogous method was applied in \cite{konmod1,konmod2}).

It is interesting that in the framework of this formulation one can avoid
complicated canonical 2nd-class constraints, usually present in canonical
LF formalism in continuous space. This greatly simplifies canonical
quantization
procedure. However the absence of explicit Lorentz invariance of
the regularization scheme makes the investigation  of the connection with
conventional Lorentz-covariant  formulation difficult. In particular,
there is a problem of the description of quantum vacuum  as common lowest
eigen state of both operators $P_-$ and $P_+$.

\subsection{Gauge-invariant action on the transverse lattice}

At first we introduce particular ultraviolet regularization via
a lattice in transverse coordinates $x^1, x^2$ and choose variables
so as to have the action, which is polynomial in these variables
\cite{barpir1,barpir2}. Furthermore, we use the described
in the introduction gauge-invariant regularization
(b) (see equation (\ref{1.2})) of
singularities at $p_- \to 0$  and gauge-invariant ultraviolet
cutoff in modes of covariant derivative $D_-$
(then ultraviolet regularization of the
theory is complete). For simplicity further we consider again
the $U(N)$ theory of pure gauge fields because this example
is technically more simple than the $SU(N)$ theory.

The components of gauge field along continuous coordinates $x^+$, $x^-$
can be taken without a modification and related to the sites of the
lattice. Transverse components are described by complex $N\times N$
matrices $M_k(x)$,
$k=1,2$. Each matrix $M_k(x)$ is related to the link
directed from the site $x-e_k$ to the site $x$.
The transverse vector $e_k$ connects two neighbouring sites on the lattice
being directed along the positive axis $x^k$ ($|e_k|\equiv a$),
see fig.~6.
 \begin{figure}[ht]
\vskip 8mm
\hskip 13mm\input fig6.pic
\caption{}
\end{figure}
The matrix $M_k^+(x)$ is related to the same link but
with opposite direction, see fig.~7.
 \begin{figure}[ht]
\vskip 8mm
\hskip 13mm\input fig7.pic
\caption{}
\end{figure}
In the following the usual rule of summation over repeated
indices is not used for the index $k$.
If this summation is necessary, the symbol of a sum is indicated.

The elements of these matrices are considered as independent variables.
This makes the action polynomial.
For any closed directed loop in the lattice we can construct the trace of
the product of  matrices $M_k(x)$ sitting on the links
in the loop and order these matrices from the right
to the left along this loop. For example the expression
 \disn{3}{
 {\rm Tr\;}\left\{M_2(x)M_1(x-e_{2})M_2^+(x-e_{1})M_1^+(x)\right\}
 \nom}
is related to the loop shown in fig.~8.
 \begin{figure}[ht]
\vskip 8mm
\hskip 13mm\input fig8.pic
\caption{}
\end{figure}

It should be noticed that a product of matrices
related to closed loop, consisting of one and the
same link passed in both directions, is not identically unity because the
matrices $M_k$ are not unitary (see, for example, fig.~9).
 \begin{figure}[ht]
\vskip 8mm
\hskip 13mm\input fig9.pic
\caption{}
\end{figure}

The unitary matrices $U(x)$ of gauge transformations act on the $M$
and $M^+$ in the following way:
 \disn{5.1a}{
M_k(x)\to M'_k(x)=U(x)M_k(x)U^+(x-e_{k}),
\nom}
 \disn{1b}{
M_k^+(x)\to M'^+_k(x)=U(x-e_{k})M_k^+(x)U^+(x).
\nom}
A trace of the product of the matrices, related to closed loop along
lattice links, is invariant with respect to
these transformations.
To relate the matrices $M_k$ with usual gauge fields of continuum theory
let us write these matrices in the following form:
 \disn{5.2a}{
M_k(x)=I+gaB_k(x)+igaA_k(x),\qquad B_k^+=B_k,\quad A_k^+=A_k.
\nom}

Then in the $a\to 0$ limit the fields $A_k(x)$ coincide with transverse gauge
field components, and the $B_k(x)$  turn out to be extra (nonphysical) fields
which should be switched off in the limit. Below we show how to get this.

The analog of the field strength
 \disn{5.2b}{
 F_{\mu\nu}=\dd_{\mu}A_{\nu}-\dd_{\nu}A_{\mu}-ig[A_{\mu},A_{\nu}],
 \nom}
multiplied by $i$, can be defined as follows:
 \disn{5.n1}{
G_{+-}=iF_{+-},\qquad F_{+-}(x)=\dd_+A_-(x)-\dd_-A_+(x)-ig[A_+(x),A_-(x)],\no
G_{\pm,k}(x)=\frac{1}{ga}\lks \dd_{\pm}M_k(x)-ig\ls A_{\pm}(x)M_k(x)-
M_k(x)A_{\pm}(x-e_k)\rs\rks,\no
G_{12}(x)=-\frac{1}{ga^2}\lks M_1(x)M_2(x-e_1)-M_2(x)M_1(x-e_2)\rks.
\nom}
 Under gauge transformation these quantities transform as follows:
 \disn{5.n2}{
G_{+-}(x)\to G'_{+-}(x)=U(x)G_{+-}(x)U^+(x),\no
G_{\pm,k}(x)\to G'_{\pm,k}(x)=U(x)G_{\pm,k}(x)U^+(x-e_k),\no
G_{12}(x)\to G'_{12}(x)=U(x)G_{12}(x)U^+(x-e_1-e_2).
\nom}

We choose a simplest form of the action having correct naive continuum
limit:
 \disn{5.n3}{
S=a^2\sum_{x^\p}\int\! dx^+\!\int\limits^L_{-L}\! dx^-\;{\rm Tr}
\lks G^+_{+-}G_{+-}+\sum_k\ls G^+_{+k}G_{-k}+
G^+_{-k}G_{+k}\rs -G^+_{12}G_{12}\rks+S_m,
\nom}
where the additional term $S_m$ gives an infinite mass to extra fields $B_k$
in the $a\to 0$ limit:
 \disn{5.n4}{
S_m=-\frac{m^2(a)}{4g^2}\sum_{x_\p}\int dx^+
\int\limits^L_{-L}dx^-\sum_k{\rm Tr}\lks\ls M_k^+(x)M_k(x)-I\rs^2\rks
\str{a\to 0}\nolr
\str{a\to 0} -m^2(a)\int d^2x^\p\int dx^+
\int\limits^L_{-L}dx^-\sum_k{\rm Tr} \ls B^2_k\rs,\qquad m(a)\str{a\to 0}\infty.
\nom}
It is supposed that this leads to necessary decoupling of the fields $B_k$.

\subsection{Canonical quantization on the Light Front}

   Let us fix the gauge as follows:
 \disn{5.n5}{
\dd_-A_-=0,\qquad A_-^{ij}(x)=\de^{ij}v^j(x^\p,x^+).
\nom}
For simplicity below we denote the argument of quantities,
not depending on the $x^-$, again by  $x$.
Let us remark that starting with arbitrary field $A_\mu$,
periodic in $x^-$, it is not possible to take
zero modes of the $A_-$ equal to
zero without a violation of the periodicity.
But it is possible to make the $A_-$
diagonal as in the equation (\ref{5.n5}) \cite{nov1,nov2,nov2a,nov3}.

Then the action (\ref{5.n3}) can be written in the  form:
 \disn{5.n6}{
S=a^2\sum_{x^\p}\int dx^+\int\limits^L_{-L}dx^-\Biggl\{
\sum_i\lks 2F^{ii}_{+-}(x)\dd_+ v^i(x)\rks+\nolr+
\frac{1}{(ga)^2}\sum_{i,j}\sum_k\lks D_-{M_k^{ij}}^+(x)\dd_+M_k^{ij}(x)
+h.c.\rks+
\sum_{i,j} A_+^{ij}(x)Q^{ji}(x)-{\cal H}(x)\Biggr\},
\nom}
where
 \disn{5.n7}{
D_-M_k^{ij}(x)\equiv \ls \dd_--igv^i(x)+igv^j(x-e_k)\rs M^{ij}_k(x),\no
D_-{M_k^{ij}}^+(x)\equiv \ls \dd_-+igv^i(x)-igv^j(x-e_k)\rs {M^{ij}_k}^+(x),\no
D_-F_{+-}^{ij}(x)\equiv \ls \dd_--igv^i(x)+igv^j(x)\rs F^{ij}_{+-}(x),
\nom}
the $A_+^{ij}(x)$ play the role of Lagrange multipliers,
 \disn{5.n8}{
Q^{ji}(x)\equiv 2D_-F_{+-}^{ji}(x)+\nol
+\frac{i}{ga^2}\sum_{j'}\sum_k
\biggl[ {M^{ij'}_k}^+(x) D_-M_k^{jj'}(x)-
{M_k^{j'j}}^+(x+e_k)D_-M_k^{j'i}(x+e_k)-\nor-
\ls D_- {M^{ij'}_k}^+(x)\rs M_k^{jj'}(x)+
\ls D_- {M_k^{j'j}}^+(x+e_k)\rs M_k^{j'i}(x+e_k)
\biggr]=0,
\nom}
are gauge constraints and
 \disn{5.n81}{
{\cal H}=\sum_{ij}\ls {F_{+-}^{ij}}^+F_{+-}^{ij} +
{G_{12}^{ij}}^+G_{12}^{ij}\rs+{\cal H}_m
\nom}
is Hamiltonian density. The term ${\cal H}_m$ can be obtained from the
expression (\ref{5.n4}) in standard way.

The constraints  can be resolved explicitly by expressing the $F_{+-}^{ij}$
in terms of other variables, but
zero mode components $F^{ii}_{+-(0)}$
 can not be found from constraint equations and play the role of independent
canonical variables. Zero modes
$Q^{ii}_{(0)}(x^{\p},x^+)$ of the constraints
remain unresolved and
 are imposed as  conditions on physical states:
 \disn{5.4.9}{
Q^{ii}_{(0)}(x^\p,x^+)\left|\Psi_{phys}\right>=0.
\nom}
In order to find complete set of independent canonical variables we write
Fourier transformation in $x^-$ of fields $M_k^{ij}(x)$ in the following
form:
 \disn{5.4.10}{
M^{ij}_k(x)=\frac{g}{\sqrt{4L}}\sum_{n=-\infty }^{\infty }
\left\{ \Theta \ls p_n+gv^i(x)-
gv^j(x-e_k)\rs M^{ij}_{nk}(x^\p,x^+)+\right.\nol
\left. +\Theta \ls -p_n-gv^i(x)+
gv^j(x-e_k)\rs {M^{ij}_{nk}}^+(x^\p,x^+) \right \}\times\nor
\times \left | p_n+gv^i(x)-
gv^j(x-e_k) \right |^{-1/2} e^{-ip_nx^-},
\nom}
where
 \disn{5.n9}{
\Theta (p) = \cases{
$1$, & $p > 0$ \cr
$0$, & $p < 0$ \cr},
\qquad p_n=\frac{\pi}{L}\,n, \quad n\epsilon Z.
\nom}
Due to the gauge (\ref{5.n5})
this Fourier transformation coincides with the expansion in
eigen modes of the operator $D_-$. Therefore the ultraviolet cutoff in
these modes, which we will apply, reduces to the following condition on the
number of terms in the sum (\ref{5.4.10}):
 \disn{5.15a}{
 |p_n + gv^i(x) - gv^j(x-e_k) | \le \frac{\pi}{L}\,\bar n,
 \nom}
where the $\bar n$ is integer parameter of ultraviolet cutoff. Let us stress
that this regularization is gauge invariant.

The action can be rewritten in the following form (up to nonessential surface
terms):
 \disn{5.4.11}{
S=a^2\sum_{x^{\p}} \int dx^+ \left\{ \sum_i 4LF_{+-(0)}^{ii}
\dd_+v^i+\right.\nolr
\left. +\frac{i}{a^2}\sum_{i,j}\sum_k{\sum_n}'
{M^{ij}_{nk}}^+\dd_+M^{ij}_{nk}+2L \sum_i
A_{+(0)}^{ii}Q_{(0)}^{ii}-\tilde {\cal H}(x) \right\},
\nom}
where the $\sum'_n $ means that the sum is cut off by the condition
(\ref{5.15a}), and
the $\tilde{\cal H}$ is obtained from the ${\cal H}$ via
the substitution of the expression
 \disn{5.n91}{
F_{+-}^{ij}=\ls F_{+-}^{ij}-\de^{ij}F_{+-(0)}^{ii}\rs+
\de^{ij}F_{+-(0)}^{ii},
\nom}
where the $F_{+-}^{ij}-\de^{ij}F_{+-(0)}^{ii}$ are to be
written in terms of the
$M^{ij}_{nk}$, ${M^{ij}_{nk}}^+$, $v^i$ by solving the constraints
(\ref{5.n8}) and using the equation (\ref{5.4.10}).
The $F_{+-(0)}^{ii}$ remain independent. The $G_{12}^{ij}$ are also to be
expressed in terms of the $M^{ij}_{nk}$, ${M^{ij}_{nk}}^+$, $v^i$
via the equations (\ref{5.n1}), (\ref{5.4.10}).

We have the following set of canonically conjugated pairs of independent
variables:
 \disn{5.4.13}{
\left \{ v^i,\quad \Pi^i=4La^2 F_{+-(0)}^{ii}\right \},\qquad
\left \{ M^{ij}_{nk},\quad i{M^{ij}_{nk}}^+ \right \}.
\nom}
In  quantum theory these variables become operators which satisfy
usual canonical commutation relations:
 \disn{5.n10}{
[v^i(x),\Pi^j(x')]_{x^+=0}=i\de^{ij}\de_{x^\p,x'^\p},\no
[M_{nk}^{ij}(x),{M_{n'k'}^{i'j'}}^+(x')]_{x^+=0}=
\de^{ii'}\de^{jj'}\de_{nn'}\de_{kk'}\de_{x^\p,x'^\p};
\nom}
the other commutators being equal to zero.

Let us remark that the condition (\ref{5.n5})
does not fix the gauge completely.
In particular, discrete group of gauge transformations,
depending on the $x^-$, of the form
 \disn{5.n10.1}{
 U_n^{ij}(x)=\de^{ij}\exp\left\{ i\frac{\pi}{L}n^i(x^{\p})x^-\right\},
 \nom}
where $n^i(x^{\p})$ are integers,
remains, and, of course, transformations, not depending on the $x^-$,
are allowed. Under the
transformations (\ref{5.n10.1}) canonical variables change as follows:
 \disn{5.n10.2}{
 v^i(x) \longrightarrow v^i(x)-\frac{\pi}{gL}\;n^i(x^{\p}),\qquad
 \Pi^i \longrightarrow \Pi^i,\no
 M_{nk}^{ij}(x^{\p}) \longrightarrow M_{n'k}^{ij}(x^{\p}),\quad
 n'=n+ n^i(x^{\p}) - n^j(x^{\p}- e_k).
 \nom}

Let us write the expression for quantum operators
$Q_{(0)}^{ii}(x^{\p},x^+)$,
which define the physical subspace of states.  We fix the order of the
operators in such a way as to relate with classical expression
$G_{\mu\nu}^{\;+}G^{\,\mu\nu}$
a quantum one of the form:
 \disn{5.n10.2a}{
 \frac{1}{2}\ls G_{\mu\nu}^{\;+}G^{\,\mu\nu} + G^{\,\mu\nu}G_{\mu\nu}^{\;+}\rs.
 \nom}
We remark that other choices of the ordering do not admit reasonable
vacuum solution.
Then the  operators $Q_{(0)}^{ii}(x^{\p},x^+)$
have the following form in terms of
canonical variables:
 \disn{5.4.16}{
2LQ^{ii}_{(0)}(x^\p,x^+)
=-\frac{g}{4a^2}
\sum_j\sum_k{\sum_n}'
\lks \e \ls p_n+gv^j(x+e_k)-gv^i(x)\rs\times\right.\hfill\nol
\times\ls
{M^{ji}_{nk}}^+(x+e_k)M^{ji}_{nk}(x+e_k)+
M^{ji}_{nk}(x+e_k){M^{ji}_{nk}}^+(x+e_k)\rs-\nor
\hfill\left.
-\e\ls p_n+gv^i(x)-gv^j(x-e_k)\rs\ls
{M^{ij}_{nk}}^+(x)M^{ij}_{nk}(x)+
M^{ij}_{nk}(x){M^{ij}_{nk}}^+(x)\rs\rks,
\nom}
where
 \disn{5.n11}{
\e(p) = \cases{
$1$, & $p > 0$ \cr
$-1$, & $p < 0$ \cr}.
\nom}

One can easily construct canonical operator of translations in the $x^-$:
 \disn{5.4.14}{
P_-^{can}=\frac{1}{2}\sum_{x^{\p}} \sum_{i,j}\sum_k{\sum_n}'
p_n\,\, \e \ls p_n+gv^i(x)-gv^j(x-e_k) \rs\times\hfill\nolr
\hfill\times\ls
{M_{nk}^{ij}}^+(x)M_{nk}^{ij}(x)+M_{nk}^{ij}(x){M_{nk}^{ij}}^+(x)\rs.
 \nom}

This expression differs from the physical gauge invariant momentum
operator $P_-$ by a term proportional to the constraint. The operator $P_-$ is
 \disn{5.4.15}{
P_-=\frac{a^2}{2}\sum_{x^{\p}} \sum_k\; \int\limits_{-L}^L dx^-\, {\rm Tr}
\ls G^+_{-k}G_{-k}+G_{-k}G^+_{-k}\rs=
P_-^{can.} + 4La^2  \sum_{x^{\p}} \sum_i
v^iQ^{ii}_{(0)}=\no
=\frac{1}{2}\sum_{x^{\p}} \sum_{i,j}\sum_k{\sum_n}'
\left| p_n+gv^i(x)-gv^j(x-e_k) \right|
\times\nolr
\times
\ls {M^{ij}_{nk}}^+(x)M^{ij}_{nk}(x)+M^{ij}_{nk}(x){M^{ij}_{nk}}^+(x)\rs
=\no
=\sum_{x^{\p}} \sum_{i,j}\sum_k{\sum_n}'
\left| p_n+gv^i(x)-gv^j(x-e_k) \right|\ls
{M^{ij}_{nk}}^+(x)M^{ij}_{nk}(x)+\frac{1}{2}\rs.
\nom}

Let us choose a representation of the state space, in which the variables
$v^i(x)$ are the multiplication operators. The states  are described in this
representation by normalizable functionals $F[v]$ of classical functions $v^i(x)$
(in fact by functions, depending on the values of the $v^i$ in different points
$x^\p$ due to the discreteness of these $x^\p$). One can define
full space of states as direct product of Fock space, in which the
${M^{ij}_{nk}}^+(x)$ and $M^{ij}_{nk}(x)$
play the role of creation and annihilation operators, and
the space of functionals $F[v]$.
Let us call $M$-vacuum the states of the form
$|0\ra\cdot F[v]$, where the $|0\ra$ satisfies the condition
 \disn{5.n12}{
 M^{ij}_{nk}(x)|0\ra=0,
 \nom}
and the $F$ is some functional. Arbitrary state can be represented in the form
of linear combination of vectors $|\{m\};F\ra$ of the form
 \disn{5.4.17}{
 \prod_{x^{\p}}\prod_{i,j}\prod_k{\prod_n}'
 \ls {M^{ij}_{nk}}^+(x)\rs^{m^{ij}_{nk}(x)}|0\ra\cdot F[v]
 \nom}
with different nonnegative integer functions  $m^{ij}_{nk}(x^\p)$
and functionals $F$. One can
define orthonormalized set of such functionals if necessary.

One can see from (\ref{5.4.15}) that the state, corresponding to the
absolute minimum of the $P_-$  must satisfy the conditions (\ref{5.n12}),
i.e. to be a $M$-vacuum.
The value of the $P_-$ in this state can be written in the form
 \disn{5.4.25}{
\langle\left. 0;F\right|P_-\left|\;0;F\right.\ra=\nolr
=\frac{1}{2}\int \prod_{x^{\p}}\prod_i dv^i(x)
\sum_{x^{\p}}\sum_{i,j}\sum_k{\sum_n}'
\left|p_n+g v^i(x)-g v^j(x-e_k)\right| |F[v]|^2.
\nom}
Remind that the ${\sum\limits_n}'$
denotes the sum in $n$ limited by the condition
(\ref{5.15a}).
If one uses this condition and shifts the index $n$ in
these sums by integer part of the quantities
$(gL(v^i(x)-v^j(x-e_k))/\pi)$, one sees that the
dependence on the $v^i$ cancels in sums over $n$
and that the expression (\ref{5.4.25})
does not depend on the $F[v]$ if it is normalized to unity.
Thus the momentum
$P_-$ has the minimum in all $M$-vacua. One can make the value of the $P_-$ in
these vacua equal to zero by subtracting corresponding constant from the
operator $P_-$.

Let us show that $M$-vacua are the physical states,
i.~e. satisfy the condition
(\ref{5.4.9}).
Indeed, in $M$-vacua this condition looks as follows:
 \disn{5.4.24}{
\sum_j\sum_k{\sum_n}'\lks
\e\ls p_n+g v^j(x+e_k)-g v^i(x)\rs-
\e\ls p_n+g v^i(x)-g v^j(x-e_k)\rs\rks F[v]=0
\nom}
and is satisfied for any $F[v]$, because for every link in the sum
(\ref{5.4.24}) the numbers of positive and negative values of
the $\e$-functions are equal.
For  arbitrary  basic states (\ref{5.4.17})
analogous conditions have the following form:
 \disn{5.4.24a}{
\sum_j\sum_k{\sum_n}'\lks
\e\ls p_n+g v^j(x+e_k)-g v^i(x)\rs m^{ij}_{nk}(x+e_k)-\right.\nolr
-\left.\e\ls p_n+g v^i(x)-g v^j(x-e_k)\rs m^{ij}_{nk}(x)\rks F[v]=0.
\nom}

The eigen values $p_-$ of the operator $P_-$ can be found from
the equation
 \disn{5.4.25a}{
\sum_{x^{\p}}\sum_{i,j}\sum_k{\sum_n}'
\left|p_n+g v^i(x)-g v^j(x-e_k)\right| m^{ij}_{nk}(x)F[v]=
p_- F[v]
\nom}
where the functional $F[v]$ is normalized.

To define  physical vacuum
state correctly one must consider not only states,
corresponding to the minimum of the operator $P_-$,
but also to the minimum of
the operator $P_+$. One can try to do this
via minimization  of the $P_+$
on the $M$-vacua, i.~e. on the set of states with $p_-=0$. The expression
$\langle 0|P_+|0\ra$,
where $|0\ra$ is the Fock space vacuum w.~r.~t. the $M_{nk}$, ${M_{nk}}^+$,
depends on the functions $v^i(x)$ (which enter, in particular, into
the "normal contractions"
of the operators $M_k$, $M_k^+$) and on the operators $\Pi^i$, canonically
conjugated to
the $v^i(x)$. The expectation value of this
expression is to be minimized on the set of
functionals $F[v]$.
The resulting functional $F[v]$ must decrease  in
the vicinity of those values of the $v^i(x)$, for which  the operator $D_-$
has zero eigen value, because
the Hamiltonian is singular at these values (it is seen, for example, from
the expansion (\ref{5.4.10})).

The vacuum state constructed in such a way
strongly differs from the usual vacuum in continuous space theory,
because for $M$-vacuum we get
zero expectation values of the operators $M_k$, but not of
the operators $(M_k - I)/ga = B_k + iA_k$,
related to usual gauge fields. Beside of this
the condition of the unitarity of the matrices $M_k$ in the continuum
limit (or equivalent condition of switching off the nonphysical fields $B_k$)
cannot be fulfilled in such a vacuum. This disagreement
with conventional theory is caused by the absence
of explicit Lorentz invariance in our formulation, that leads to
different quantum states, corresponding to the
minima of the operators $P_-$ and
$P_+$. It is not clear whether these states can be made coinciding at least in
the limit of continuous space. This requires further investigation.

Nevertheless the method of the quantization of gauge theories on the LF, described
here, can be useful for completely gauge-invariant formulation of some effective
models, based on analogous formalism (but
without complete gauge invariance due
to throwing out of all zero modes of fields and due to the absence of
gauge-invariant
regularization of ultraviolet divergencies in the $p_-$).
Such models are described, for example, in papers \cite{dalley1,dalley4},
where the ideas of papers \cite{pirner1,mack} were developed.

\section{Conclusion}

The LF Hamiltonian approach to Quantum Field Theory
briefly reviewed here is an attempt to apply a beautiful
idea of Fock space representation for quantum field
nonperturbatively in the framework of canonical
formulation on the LF. The problem of describing
the physical vacuum state becomes formally trivial
in this approach because such vacuum state  coincides
with mathematical vacuum of LF Fock space.

     However the breakdown of Lorentz and gauge
symmetries due to regularizations generates difficulties
in proving the equivalence of LF formalism and the usual
one in Lorentz coordinates. This problem can be solved
for nongauge theories but turns out to be very difficult for
gauge theories in special (LF) gauge which is needed
here. Nevertheless we hope that these difficulties can
be overcome by finding a modified form of canonical
LF Hamiltonian which generates the perturbation
theory equivalent to usual covariant and gauge invariant
one.

\vskip 1em
{\bf Acknowledgments.}
The work was supported by the Russian Foundation
for Basic Research,
grant no.~05-02-17477 (S.A.P. and E.V.P.).

\vskip 7mm
\setcounter{equation}{0}
\renewcommand{\theequation}{A.1.\arabic{equation}}
\section*{$\protect\vphantom{a}$\hfill Appendix 1}
{\bf Statement 1.} {\it If conditions (\ref{4.1.1}) are  satisfied,
then,  for  fixed
external momenta $p^s$ and $p^s_-\ne    0 \;\forall s$, the  equality
 \disn{4.A1}{
\lim_{\bet\to    0}\lim_{\g\to    0}
\int\prod_k  dq_+^k  \int\limits_{V_{\scriptstyle  \e}}
\prod_k dq_-^k    {{\tilde    f(Q^i,p^s)    e^{-\g\sum_i
{Q_+^i}^2-\bet
\sum_i{Q_-^i}^2} }\over{\prod_i (2Q_+^iQ_-^i-M^2_i+i\ka)}}=\nolr
=\int\prod_k dq_+^k \int\limits_{V_{\scriptstyle \e}\cap B_L}
\prod_k dq_-^k
{{\tilde f(Q^i,p^s)}\over{\prod_i (2Q_+^iQ_-^i-M^2_i+i\ka)}},
\nom}
holds while the expressions appearing in  (\ref{4.A1}) exist  and  the
integral  over  $\{q_+^k\}$  on  the  right-hand  side  is    absolutely
convergent. It is assumed  that  the  momenta  of  lines  $Q^i$ are
expressed in terms of loop momenta $q^k$, $V_{\e}$
is the domain corresponding to the presence of full lines,
type-1  lines,  and  type-2  lines  (the  definitions  are  given
following formula (\ref{4.22})), $B_L$ is the sphere of radius $L$,
where $L\ge S \: \max\limits_s |p_-^s|$, and $S$
is a number depending on the diagram structure.  }

Let us prove the statement. For each type-1 line in  (\ref{4.A1}),  we
perform the following partitioning:
 \disn{nnn9}{
\int\limits_{-\e}^{\e} dQ_-^i=\lks \int dQ_-^i+(-1)
\ls  \int\limits_{-\infty}^{-\e}
dQ_-^i + \int\limits_{\e}^{\infty} dQ_-^i \rs \rks .
\nom}
Then both sides of equation (\ref{4.A1}) become the sum of  expressions  of
the same form in which, however, the domain $V_{\e}$ corresponds to the
presence of only full and type-2  lines.  It  is  clear  that  by
proving the statement for this $V_{\e}$:  (which  is  done  below),  we
prove the original statement as well.

Let $\tilde B$ be a domain such that the surfaces on which
\hbox{$Q_-^i=0$} are not
tangent to the boundary $\tilde B$. First, we prove that in the expression
 \disn{4.A2}{
\int\prod_k dq_+^k
\int\limits_{V_{\scriptstyle \e}\cap \tilde B} \prod_k dq_-^k
{{\tilde f(Q^i,p^s)\: e^{-\bet \sum_i{Q_-^i}^2}}\over{\prod_i
(2Q_+^iQ_-^i-M^2_i+i\ka)}}
\nom}
the integral  over  $\{q_+^k\}$  is  absolutely  convergent  (here  the
integral over $\{q_-^k\}$ is finite because x$\ka>0$,   $\bet>0$).
This  becomes
obvious (considering conditions (\ref{4.1.1}) and the fact that, in  type-2
lines, the momentum $Q_-^i$ is separated from zero) if  the  contours
of the integration over $\{q_-^k\}$  can be deformed in such a way  that
the momenta $Q_-^i$ of the full lines are separated from zero  by  a
finite quantity (within the domain $V_{\e}\cap \tilde  B$).
In this case, we  can
repeat the well-known Weinberg reasoning \cite{wein}.  What  can  prevent
deformation is either a "clamping" of the contour  or  the  point
$Q_-^i=0$ falling on the integration boundary.

Let us investigate the first alternative. We divide the domain of
integration over $q_+^k$  into sectors such that  the  momenta  of  all
full lines $Q_+^i$ have a constant sign  within  one  sector.  Let  us
examine one sector. We take a set of full  lines  whose  $Q_-^i$  may
simultaneously vanish. In the vicinity of  the  point  where  $Q_-^i$
from this set vanish simultaneously, we bend the contours of  the
integration over $\{q_-^k\}$ such that these contours pass  through  the
points $Q_-^i=iB^i$ and the momenta $Q_-^i$ of the type-2 lines  do  not
change. Let $B^i$ be such that $B^i  Q^i_+  \ge  0$
for the lines from  the  set
(for $Q^i_+$  from the sector under  consideration).  It  is  easy  to
check that this bending  is  possible.  (Since  the  contours  of
integration over $q_-^k$ are bent and $Q^i_-$ are expressed in terms of
$q_-^k$,
one should only check that such $b^k$ exist, where the necessary  $B^i$
are  expressed  in  the  same  way,  i.e.,  that  $B^i$  obey    the
conservation laws and flow only along the full lines). With this
bending, rather small in relation to the deviation and the  size
of the deviation region, the contours do not pass through  the
poles because, for the denominator of each line from the set  in
question, we have
 \disn{nnn10}{
\ls 2Q_+^iQ_-^i-M^2_i+i\ka\rs \to
\ls 2Q_+^i\ls Q_-^i+iB^i\rs -M^2_i+i\ka\rs , \quad Q_+^iB^i\ge 0,
\nom}
and for the other denominators, the  bending  takes  place  in  a
region separated from the point where the corresponding momenta
$Q_-^i$ are equal to $0$. Repeating the reasoning for all sets, we  can
see that there is no contour "clamping".

The other alternative is excluded by the above condition  for $\tilde B$.
To make this clear, one should introduce such coordinates $\xi^{\al}$  in
the $q^k$-space that the boundary of the domain $\tilde   B$
is  determined by the
equation $\xi^1=a=const$ and then argue  as  above  for  the
coordinates $\xi^{\al}$ with $\al\ge 2$.

After bending  the  contours,  integral  (\ref{4.A2}) is  absolutely
convergent in $q_+^k$, $q_-^k$ if the integration in $q_+^k$
is  carried  out
within the sector under consideration. On pointing out that the
result, of internal integration in (\ref{4.A2}) does not  depend  on
the  bending,  we  add  the  integrals  over  all  sectors    and
conclude that (\ref{4.A2}) converges in $\{q_+^k\}$ absolutely.

Now let us prove that if $\tilde  B$ is a quite small, finite  vicinity  of
the  point  $\{\tilde q_-^k\}$  that  lies  outside
the  sphere  $B_L$,  then
expression (\ref{4.A2}) is equal to zero. We  consider  the  momentum
$Q_-^i$ of one line. Flowing along
the diagram, it can ramify or it can merge  with  other  momenta.
Clearly, two  situations  are  possible:  either  it  flows  away
completely through  external  lines,  or,  probably,  after  long
wandering, part of it, $\tilde Q_-$, makes a  complete  loop.  The  former
situation is possible only if $|Q_-^i|\le \sum_r |p_-^r|$,
where all external
momenta leaving the diagram (but not  entering  it)  are  summed.
Obviously, $S$ can be chosen such that for  $\{q_-^k\}$
from  $\tilde B$,  a  line
exists whose momentum violates this condition.

The latter situation results in the existence of  a  loop,  where
the inequality $Q_-^i>\tilde    Q_-$
holds for all momenta of  its  lines  and
the positive direction of the momenta is along the loop. Then the
integral over $q_+^k$ of the loop in question can be interchanged with
the integrals over $\{q_-^k\}$ (because it is absolutely  and  uniformly
convergent for all $q_-^k$) and the residue formula  can  be  used  to
perform this integration. Since, for the loop  in  question,  the
momenta $Q_-^i$  of the lines of this loop are separated from zero and
are of the same sign, the result  is  zero.  This  has  a  simple
physical  meaning.  If  we  pass  to   stationary    noncovariant
perturbation theory, we find that only quanta  with  positive  $Q_-$
can exist. In this case, external particles with positive $p_-$  are
incoming and those with  negative  $p_-$  are  outgoing.  Then,  the
momentum conservation law favors  the  occurrence  of  the  first
situation.

The entire outside space for $\tilde B$ can  be  composed  of  the  above
domains $B_L$ (everything converges well
at infinity due to the  factor  $\exp(-\bet  \sum_i{Q_-^i}^2)$).
Thus,  on  the
left-hand side of (\ref{4.A1}), one  can  substitute  the  integration
domain $V_{\e}\cap B_L$ for $V_{\e}$, set the limit  in  $\g$
under  the  sign  of
integration over $\{q_+^k\}$ because of its  absolute  convergence,  and
also set the limit in $\bet$ under the integration sign  because  the
domain of the integration over $\{q_-^k\}$ is bounded. Thus, we  obtain
the right-hand side. The statement is proved.

{\bf Statement 2.} {\it If $V_{\e}$ corresponds  to  the  presence  of
type-2  lines alone, then, under the same conditions as in
Statement~1,  the equality
 \disn{nnn11}{
\int\limits_{V_{\scriptstyle \e}}\prod_k dq_-^k\int\prod_k dq_+^k
{{\tilde f(Q^i,p^s)}\over{\prod_i (2Q_+^iQ_-^i-M^2_i+i\ka)}}=\nolr
=\int\prod_k dq_+^k \int\limits_{V_{\scriptstyle \e}\cap B_L}
\prod_k dq_-^k
{{\tilde f(Q^i,p^s)}\over{\prod_i (2Q_+^iQ_-^i-M^2_i+i\ka)}}.
\nom}
is valid.  }

The proof of this statement is analogous to the  second
part of the proof of Statement~1.

\vskip 7mm
\setcounter{equation}{0}
\renewcommand{\theequation}{A.2.\arabic{equation}}
\section*{$\protect\vphantom{a}$\hfill Appendix 2}
{\bf Statement.} {\it If conditions (\ref{4.1.1})
are satisfied, the limits in  $\g$  and
$\bet$ in (\ref{4.18}) can be interchanged (in turn)  with  the  sign  of  the
integral over $\{\al_i\}$ and then with
$\tilde f\ls -i{{\dd}\over{\dd y_i}}\rs $.  }

To prove this, we  define  the  vectors
\disn{nnn12}{
\{q_+^1,q_-^1,\dots, q_+^l,q_-^l\}\equiv S,\qquad
\{Q_+^1,Q_-^1,\dots, Q_+^n,Q_-^n\}\equiv \m S+P,\no
\{y_1^+,y_1^-,\dots, y_n^+,y_n^-\}\equiv Y,
\nom}
where the vector $P$ is built only from external momenta and  $\m$  is
an $l\times    n$ matrix of rank $l$,
$\m^{2i}_{2k-1}=\m^{2i-1}_{2k}=0$,
$\m^{2i}_{2k}=\m^{2i-1}_{2k-1}$. Next,  we
introduce the following notation:
 \disn{nnn13}{
\tilde \La_i=\ls \begin{array}{cc}
\g&-i\al_i \\
-i\al_i&\bet
\end{array}\rs ,\quad
\La=diag\{\tilde \La_1,\dots,\tilde \La_n\}, \quad A=\m^t\La\m,\no
B=\m^t \La P -{1\over 2}i\m^t Y,\quad
C=-P^t\La P+iY^t P-i\sum_i\al_i M^2_i.
\nom}
Then it follows from (\ref{4.19}) that
 \disn{4.B4}{
\hat \f(\al_i,p^s,\g,\bet)=(-i)^n \tilde f\ls -i{{\dd}\over{\dd y_i}}\rs
\int d^{2l}S \; e^{-S^tAS-2B^tS+C}\Bigr|_{y_i=0}=\nolr
=(-i)^n \tilde f\ls -i{{\dd}\over{\dd y_i}}\rs
e^{B^tA^{-1}B+C}{{\pi^l}\over{\sqrt{\det A}}}\biggr|_{y_i=0}.
\nom}
The function $\tilde f$ is a polynomial and we consider each of its  terms
separately. Up to a factor, each term has
the form ${{\dd}\over{\dd y_{i_1}}}\dots {{\dd}\over{\dd  y_{i_r}}}$.
These derivatives act on $C$ and $B$. The action
on $C$ results in the constant factor $iN^tP$, the action on $B$
results in  the  factor
$-(1/2)iN^t\m  A^{-1}B$ or $-(1/4)N_1^t\m  A^{-1}\m^tN_2$
(the latter is the result of  the  action  of  two
derivatives; $N$, $N_1$, and $N_2$ are constant vectors).

It is necessary to prove the correctness of the  following  three
procedures: (i) setting the limit in $\g$ under the  integral  sign
for fixed $\bet>0$; (ii) setting the limit in $\bet$ for $\g=0$;  (iii)
setting  the  limits  in  $\g$  and  $\bet$  under   the    signs    of
differentiation with respect to $Y$. In cases  (i)  and  (ii),  one
must obtain the bounds
 \disn{4.B5}{
|\hat \f(\al_i,p^s,\g,\bet)|\le \f' (\al_i,p^s,\bet),
\nom}
 \disn{4.B6}{
|\hat \f(\al_i,p^s,0,\bet)|\le \f'' (\al_i,p^s),
\nom}
where $\f'$ and $\f''$ are functions integrable (for $\f'$ if $\bet>0$)  in
any finite domain over $\al_i$, with $\al_i\ge 0$.
Then, for case (i), we have
 \disn{nnn14}{
|\hat \f(\al_i,p^s,\g,\bet)\; e^{-\ka\sum_i \al_i}|\le
\f' (\al_i,p^s,\bet)\; e^{-\ka\sum_i \al_i},
\nom}
i.e., a limit on the integrated function arises, and,  thus,  the
limit in $\g$ can be put under the integral sign. The  situation  is
similar for case (ii).  It  is  evident  from  (\ref{4.B4})  that  the
function $\hat \f$ can be singular only if the eigenvalues of  matrix  $A$
become zero. On finding the lower bound of these eigenvalues, one
can prove through rather  long  reasoning  that  bounds  (\ref{4.B5}),
(\ref{4.B6}) exist if condition (\ref{4.1.1}) is satisfied.

After the limits in $\g$ and $\bet$ are put under the integral sign,  it
is not difficult to interchange  them  with  the  differentiation
with respect to $Y$. One need do it only for $\al_i>0$  (for  each $i$)
and, in this case, one can  show  that  the  eigenvalues  of  the
matrix $A$ are nonzero and $\hat \f$ is not singular.

\end{document}